\documentclass[AMA,STIX1COL]{WileyNJD-v2}

\usepackage{moreverb}
\usepackage{amsmath,amssymb,amsfonts}
\usepackage{graphicx}
\usepackage{textcomp}
\usepackage{hyperref}
\usepackage{times}
\usepackage{color}
\usepackage{amsthm}
\usepackage{subfig}
\usepackage{commath}
\usepackage{tikz}
\usepackage{scalerel}
\usepackage{comment}
\usepackage{bbm}
\newcommand{\R}{\mathbb{R}}
\usepackage{siunitx}
\usepackage{todonotes}
\usepackage{textcase}

\newcommand\BibTeX{{\rmfamily B\kern-.05em \textsc{i\kern-.025em b}\kern-.08em
T\kern-.1667em\lower.7ex\hbox{E}\kern-.125emX}}

\newcommand{\reals}{{\mathbb R}}
\newcommand{\leftm}{\left[\begin{array}}
	\newcommand{\rightm}{\end{array}\right]}
\newcommand{\T}{^{\top}}

\renewcommand{\cos}{\mathrm{c}}
\renewcommand{\sin}{\mathrm{s}}

\renewcommand{\emptyset}{\varnothing} % empty set
\newcommand{\Rset}[1]{\mathbb{R}^{#1}} % R^n
\def \b {\mathbf{b}}

\def \f {\mathbf{f}} 
\def \g {\mathbf{g}}

\def \n {\mathbf{n}} 
 
\def \p {\mathbf{p}} 
 
\def \r {\mathbf{r}} 
\def \s {\mathbf{s}} 
 
\def \u {\mathbf{u}} 
 
\def \w {\mathbf{w}}

\def \P {\mathbf{P}} 
 
\def \R {\mathbf{R}} 
\def \S {\mathbf{S}} 
\def\Cmc{\mathcal{C}}

\def\Smc{\mathcal{S}}

\def\Vmc{\mathcal{V}}

\newcommand{\Bsigma}{\boldsymbol{\sigma}}

\newcommand{\Bupsilon}{\boldsymbol{\upsilon}}

\newcommand{\Bomega}{\boldsymbol{\omega}}

\def \zeros			{{\mathbf{0}}} % ones vector

\articletype{Article Type}%

\received{<day> <Month>, <year>}
\revised{<day> <Month>, <year>}
\accepted{<day> <Month>, <year>}

\begin{document}

\title{Bearing-based Autonomous Communication Relay Positioning under Field-of-View Constraints}

\author[1]{Marco Fabris*}

\author[1]{Daniel Zelazo}

%\author[3]{Author Three}

\authormark{M. FABRIS \textsc{et al}}

\address[1]{\orgdiv{Faculty of Aerospace Engineering}, \orgname{Technion - Israel Institute of Technology}, \orgaddress{\state{Haifa District}, \country{Israel}}}

%\address[2]{\orgdiv{Org Division}, \orgname{Org name}, \orgaddress{\state{State name}, \country{Country name}}}

%\address[3]{\orgdiv{Org Division}, \orgname{Org name}, \orgaddress{\state{State name}, \country{Country name}}}

\corres{*Marco Fabris, Faculty of Aerospace Engineering, Technion - Israel Institute of Technology, Technion City, Haifa 3200003, Israel. \email{marco.fabris@campus.technion.ac.il}}

%\presentaddress{Present address}

\abstract[Abstract]{This paper investigates the problem of communication relay establishment for multiple agent-based mobile units using a relay vehicle. The objective is to drive autonomously the relay vehicle to attain a position for communication relay establishment while maintaining the other vehicles inside of its field-of-view. A bearing-based control law is proposed for the relay drone and designed for both single and multiple agents. We also provide a collision avoidance scheme that ensures no collisions between the relay and other agents. Numerical simulations and experimental results are reported as well to show the efficacy of the proposed approach.}


%\jnlcitation{\cname{%
%\author{Williams K.}, 
%\author{B. Hoskins}, 
%\author{R. Lee}, 
% \author{M. Fabris} and 
% \author{D. Zelazo}}, 
% \ctitle{``Bearing-based Autonomous Communication Relay Positioning under Field-of-View Constraints''}, \cjournal{Advanced Control for Applications}, \cyear{2022}.
%}

\maketitle

\footnotetext{\textbf{Abbreviations:} GPS, global positioning system}

\section{Introduction} 

With recent technological advances, the use of unmanned aerial vehicles (UAVs), or more generally \textit{drones}, has become widespread both in practical applications and in many areas of research. Due to their relatively low cost, small size and increased efficiency, one of the key areas in which drones have found themselves being investigated is in the robotics and communications community, mainly focusing on coordination of drones using relays for ensuring connectivity \cite{Yanmaz2017, Isarabhakdee2009, DeMoraes2017, LunYaoWang2020}. In addition, new research has shown that drones are able to effectively be employed as communication relays \cite{OreillyChaffey2015,CetinZagli2012} between \textit{ground units}, allowing them to overcome communication difficulties caused by obstacles or terrain features, such as buildings or mountainous terrain. The approach in these works often assumes a centralized level of coordination \cite{WuGaoWan2020}, known locations of ground units \cite{Ono2016,Chen2018-OptPlacementRelay}, or focuses on a small finite subset of connectivity topologies within decentralized frameworks \cite{KimLadoszOh2020,RenAtkins2007}. % or focuses on higher-level mission planning strategies. 

Part of the latest research has been dedicated to positioning a relay drone when the initial ground unit locations are unknown and for mobile ground units \cite{Chamseddine2017-CommRelayMultiground}; nevertheless, further limitations arise as the relay drone is assumed to be able to perform tracking of ground units only from a fixed altitude \cite{I-HuaiLu12017-DevInstRelay}. Indeed, in many applications, such as those for surveillance tasks on mountainous terrain \cite{BurdakovDoherty2009}, we find situations in which a relay drone should ideally not be set at a constant {vantage point}, but rather should self-adjust to any required position to communicate with other agents in the network. To solve for this, we devise an approach based on \textit{multi-agent formation control} \cite{OH2015424,ahn2020formation,OhAhn2010,FabrisCenedeseHauser2019}. At its most fundamental level, formation control involves the coordination of a team of agents to achieve some spatial formation shape. From a control systems perspective, the challenge in formation control is to find distributed strategies, e.g. \cite{LesserOrtizJrTambe2012,BishopDeghatAnderson2015} for the control and estimation of \textit{multi-agent systems} (MASs) \cite{ChenRen2019,Mesbahi2010} that achieve a desired formation with guarantees on certain properties such as stability and performance. 

Remarkably, the sensing and communication capabilities in a MAS largely influence the resulting control strategies that can be employed. Indeed, if all agents had access to accurate and reliable global state information (such as from GPS) with the ability to broadcast their state to the entire ensemble, then strategies for achieving formation control could be approached using more traditional methods from modern control theory (see also the study in \cite{GriffinFierroPalunko2012}, coping with GPS-denied environments). Nonetheless, the measurements available for each agent to achieve the task of formation control are often inherently local in nature and nonlinear functions of the agent states. Common examples include range sensors for measuring the distance between agents and bearing sensors for measuring the bearing angle from the body frame of one agent to another agent. For such sensing mediums, the combinatorial discipline known as \emph{rigidity theory} \cite{de2019formation} has emerged as the correct framework for studying these formation control problems.

Rigidity is a combinatorial theory for characterizing the ``stiffness'' or ``flexibility'' of structures formed by rigid bodies connected by flexible linkages or hinges. In \cite{Krick2009IJC}, it was shown that formation stabilization using distance measurements can be achieved only if rigidity of the formation is maintained. Formation rigidity also provides a necessary condition for estimating relative positions using only relative distance measurements \cite{2006-AspEreGolMorWhiYanAndBel,2010-CalCarWei}. Similar to distance-based rigidity theory, a novel extension based on bearing measurements has been lately developed, known as \emph{bearing rigidity} \cite{MichielettoCenedeseFranchi2016,michieletto2020unified,ZhaoZelazo2019} (sometimes referred to as parallel rigidity).  Whereas rigidity theory is useful for maintaining formations with fixed distances between neighboring agents, bearing rigidity focuses on maintaining formation shapes, that is it attempts to keep the bearing vector between neighboring agents constant. Bearing rigidity was used in \cite{Eren2012, Franchi2012a, TrinhMukherjeeZelazo2018,KoHoangAhn2020} for deriving distributed control laws for controlling formations with bearing measurements. In \cite{2009-ShaFidAnd, 2003-EreWhiMorBelAnd, Eren2007, 2006-AspEreGolMorWhiYanAndBel}, it was employed for the localization problem in robotic networks using bearing measurements. Bearing rigidity has also proven useful for stabilization of formations using direction-only constraints \cite{2011-BisShaAnd, 2003-EreWhiMorBelAnd}.

To the author's best knowledge, most existing works assume a static information exchange network and do not incorporate real-world constraints such as sensing and communication range and field-of-view constraints  \cite{HoaNguyenHobart2002,JohansenZolichHansen2014,KimLadoszOh2020}. However, these real-world constraints introduce state-dependent nonlinearities to the information exchange problem which can be difficult to solve.  This aspect was partially addressed for the formation control problem in \cite{Zelazo2013a_J, FrankZelazo2018}.  

\textbf{Contributions}: In this paper, we study the feasibility of using a consolidated bearing-based-formation-control approach  to drive an autonomous \textit{relay vehicle} (RV) on a two-dimensional scenario in order to maintain relay capability between multiple aerial and/or ground vehicles addressed as the \textit{agents} (that is, a robotic MAS), while keeping said agents inside a given \textit{field-of-view} (FoV). The reasons for this are disparate, e.g. preservation of connectivity for the underlying network of agents \cite{SantilliMukherjeeGasparri2019} or surveillance of inaccessible regions \cite{KumarGhoshSinghal2020}.

Our major contribution rests upon an innovative method leveraging bearing-based formation control in which the need for both a centralized level of coordination and distance measurements is completely removed, providing advantages such as mitigating the effect of a single point of failure and, {more generally, expanding the most recent and cutting edge findings in this research front (see, e.g.,~\cite{LiZhaoMiaoZhang2021})}. More precisely, the objective is to drive the RV to a position for relay establishment in a planar environment while maintaining the other vehicles inside of its FoV and avoiding collisions. To this purpose, a bearing-based control law is proposed for the RV guidance and designed to track both single and multiple vehicles. In particular, by tuning the relay control gain above a certain threshold derived analytically in closed form, it is shown that a MAS can be tracked at the best relay capabilities over time. It is then guaranteed that the agents on the verge to exit the RV's FoV are continuously tracked over time, implying that the whole considered MAS is kept in the RV's FoV, whenever a stable formation protocol is employed to govern it. Furthermore, the implementation of a specific collision avoidance strategy for this aim is also developed, representing our second main contribution. Such theoretical advances are finally validated by numerical simulations.

\textbf{Paper organization}: The remainder of this paper is arranged as follows. In Sec. \ref{sec:preliminaries}, we introduce mathematical preliminaries and examine more deeply the concepts of bearing-based formation control. Sec. \ref{sec:systemmodelandcontrolstrategy} describes the system model adopted and defines our proposed control strategy, analyzing its working principles and implementing an effective collision avoidance strategy. Sec. \ref{sec:simulations} is devoted to the results of our numerical simulations. Finally, Sec. \ref{sec:conclusions} briefly concludes our work, discussing future directions, and App. \ref{sec:appendix} contains the appendix.

\textbf{Basic notation}: Hereafter, symbol $\reals$ denotes the set of real numbers. Letter $t$ addresses continuous time instants. We indicate with $\mathbf{I}_m \in \reals^{m \times m}$ and $\zeros_{}$ the identity matrix of dimension $m$ and the null vector, respectively. Moreover, symbols $\T$ and $|A|$ denote the transpose operator and the cardinality of set $A$, respectively. Let $\mathbf{w} \in \reals^{m}$ be a vector, then $\left\|\mathbf{w} \right\|$ denotes its Euclidean norm, i.e. $\left\|\mathbf{w} \right\|^{2} = \mathbf{w}\T \mathbf{w}$. Given an angle $\alpha \in [0,2\pi)$, we use the short notation $\cos_{\alpha} = \mathrm{cos}(\alpha)$ and $\sin_{\alpha} = \mathrm{sin}(\alpha)$. For $\alpha \in [-1,1]$, $\arccos(\alpha)$ and $\arcsin(\alpha)$ indicate the inverse cosine and sine functions of $\alpha$. Lastly, $\mathrm{sign}(\alpha)$, with $\alpha \in \reals$, addresses the sign function that returns $1$ if $\alpha > 0$; $-1$ if $\alpha <0$; $0$ if $\alpha = 0$ and $[\alpha]_{+} = \alpha$, if $\alpha \geq 0$; $[\alpha]_{+} = 0$, otherwise.

%%%%%%%%%%%%%%%%%%%%%%%%%%%%%%%
%%%%%%%%%%%%%%%%%%%%%%%%%%%%%%%
%%%%%%%%%%%%%%%%%%%%%%%%%%%%%%%

\section{Preliminaries} \label{sec:preliminaries}
Several tools from bearing-based formation control may assume a crucial role while identifying an effective strategy for autonomous communication relay positioning under FoV constraints. 
In this study, we consider a team of $n$ agents and denote by $\mathbf{p}_i = \mathbf{p}_i(t) \in \reals^d$ the position of agent $i$ in a $d$-dimensional Euclidean space, so that $\mathbf{p}_i= \begin{bmatrix}
	p_{1,i} & \cdots & p_{d,i}
\end{bmatrix}\T$. This choice is motivated by the fact that, frequently, an abstraction made in the formation control literature is to model each vehicle as a simple kinematic point mass \cite{Krick2009IJC, Marshall2004, ZhaoSun2017}. Moreover,  
 in practical applications, autonomous vehicles are often modeled in 2D and 3D spaces ($d=2,3$). Here, $d=2$ is set, as we deal with mobile robots deployed on planar environments.
The spatial configuration of all the agents is then denoted by the stacked vector $\mathbf{p} = \begin{bmatrix}
	\mathbf{p}_{1}\T & \cdots & \mathbf{p}_{n}\T
\end{bmatrix} \T$.  
Generally, each vehicle is able to sense certain quantities that are a function of their relative states, such as the distance between vehicle $i$ and $j$ defined as $d_{ij} = \|\mathbf{p}_j - \mathbf{p}_i\|$, (e.g., in distance-based and displacement-based formation control \cite{FabrisCenedeseHauser2019}), 
or the bearing between vehicle $i$ to vehicle $j$, denoted by $\mathbf{g}_{ij}$ and defined by the unit vector
\begin{equation}\label{eq:bearingvectordef}
	\mathbf{g}_{ij} = \dfrac{\mathbf{p}_j-\mathbf{p}_i}{\|\mathbf{p}_j - \mathbf{p}_i\|} ,
\end{equation}
 (e.g., in bearing-based formation control \cite{zhao2014TACBearing}). In the following, bearing vectors of the same type of \eqref{eq:bearingvectordef} are going to be used for strict guidance purposes; whereas, short-range distance measurements are going to be employed in collision avoidance only.

In addition, the sensing and communication topology of a multi-vehicle system is here described by a graph \cite{BookGodsilGraph}.  
A graph, here denoted by $\mathcal{G} = (\mathcal{V},\mathcal{E})$, is defined by a set of nodes, $\mathcal{V}= \{1,\ldots,n\}$, and a set of edges, $\mathcal{E} \subseteq \mathcal{V} \times \mathcal{V} $, describing the incidence relationship between nodes.  Thus, agent $i$ is associated to the node $i \in \mathcal{V}$ in the graph %(simply denoted by $i$ in the sequel, with a little abuse of notation), and vehicle $i$ 
and has access to a relative measurement with vehicle $j$ if and only if $e_{ij} := (i,j) \in \mathcal{E}$. In the sequel, we indicate the neighborhood of node $v_{i}$ as the set $\mathcal{N}_{i} = \{j ~|~ (i, j) \in \mathcal{E}\}$. %The combination of the interaction graph and the spatial configuration of the vehicle team, $(\mathcal{G},\mathbf{p})$, is referred to as a \emph{framework} \cite{Asimow1978Rigidty}.

%With the above set-up, the formation control problem can be formally defined. 

With the above set-up, the foundations of the control strategy we are going to pursue assume that the robots are able to measure the bearing angle to neighboring agents and the formation is also specified by bearing measurements.  In this direction, we recall that a desired formation can be characterized by set $\mathcal{F}(\mathbf{p}) = \{ \mathbf{p} \in \reals^{nd} \, | \, \mathbf{g}_{ij} = \mathbf{g}^{*}_{ij} , \, \forall e_{ij} \in \mathcal{E}\}$, wherein quantities $\mathbf{g}^{*}_{ij}$ represents the $(i,j)$-th desired bearing to be achieved. Note that set $\mathcal{F}(\mathbf{p})$ is specified by bearings \eqref{eq:bearingvectordef} only and,
as shown in \cite{zhao2014TACBearing}, its use naturally leads to the adoption of the following gradient-like formation control law
\begin{equation}\label{bearing_control_law}
	\dot{\mathbf{p}}_i = - \sum_{j \in \mathcal{N}_{i}} \mathbf{P}_{\mathbf{g}_{ij}} \mathbf{g}_{ij}^*, \quad i \in \{1, \ldots, n\} ,
\end{equation}
where 
\begin{equation}\label{eq:def_proj_matr_bearing}
	\mathbf{P}_{\mathbf{g}_{ij}} = \mathbf{I}_d - \mathbf{g}_{ij}\mathbf{g}_{ij}\T \in \reals^{d \times d}
\end{equation}
is an orthonormal projection operator. Observe that control law \eqref{bearing_control_law} is distributed, as each neighbor only relies on the measured bearing to its neighbors and the desired bearing angle.  The control also has a geometric interpretation, since each term $\mathbf{P}_{\mathbf{g}_{ij}} \mathbf{g}_{ij}^*$ in \eqref{bearing_control_law} is orthogonal to $\mathbf{g}_{ij}$, for all $t \geq 0$, implying that the bearing-based control law \eqref{bearing_control_law} attempts to reduce the bearing error between agents $i$ and $j$. 

In the next section, the above preliminaries provide the key in the development of a control law that allows an additional robotic entity, the so-called RV, to track a MAS having its own dynamics. % modeled as in \eqref{integrator}.

%%%%%%%%%%%%%%%%%%%%%%%%%%%%%%%
%%%%%%%%%%%%%%%%%%%%%%%%%%%%%%%
%%%%%%%%%%%%%%%%%%%%%%%%%%%%%%%

\section{Autonomous Relay Tracking}\label{sec:systemmodelandcontrolstrategy}
We consider a distinct variation of the bearing-only formation control problem wherein one designated agent, i.e. the RV, is tasked with maintaining a line-of-sight measurement to one or more other agents only.  Each agent is tasked with its own individual mission and includes way point tracking or variations of the coverage control problem (see \cite{FabrisCenedeseCoverage2019,GALCERAN20131258}). %,1284411  
Here, the agents may not have any coordination constraints, in principle, meaning that they are not required to perform any collaborative tasks such as formation control.  It is simply assumed each agent has its own task to perform.  Moreover, the RV does not have any information on the trajectories of the independent agents except some basic dynamic constraints of the agents, such as their maximum speed.  Thus, the proposed control strategy for the relay is based only on \emph{sensed} information and relative state information between itself and the agents.
%\todo[inline]{perhaps we refer to the "other UAvs " simply as agents - everywhere.  This way we don't restrict ourselves to only uavs, but any autonomous systems (with applicatinos in land, sea, and space)}

\subsection{System model and problem statement}
We assume that the agents and RV have full knowledge of a global inertial frame and operate on a planar environment ($d=2$).  
The relay dynamics is modeled as a first-order integrator, namely
\begin{equation}\label{integrator_dynamics}
	\dot{\mathbf{p}}_r(t) = \mathbf{u}_r(t),
\end{equation}
where $\mathbf{p}_r(t) \in \mathbb{R}^{d}$ is the relay position and $\mathbf{u}_r(t) \in \mathbb{R}^{d}$ is the relay control in velocity; whereas, the agents' dynamics is solely characterized by the presence of a global upper bound for the $i$-th agent velocity $\dot{\p}_{i}$, $i= 1,\ldots,n$. Formally:
\begin{assumption}\label{asm:max_speed}
	The speed of each agent is upper bounded by the constant $v_M > 0$ (known by the RV), such that 
	\begin{equation*}
		\left\| \dot{\p}_{i}(t) \right\| \leq v_{M}, \quad \forall i\in \Vmc, \forall t\geq 0.
	\end{equation*}
\end{assumption}
\begin{figure}[!b]
	\centering
	% trim=left bottom right top
	\hspace{-1.2cm}
	\includegraphics[scale=0.45,trim=8cm 6.7cm 7cm 0.2cm, clip]{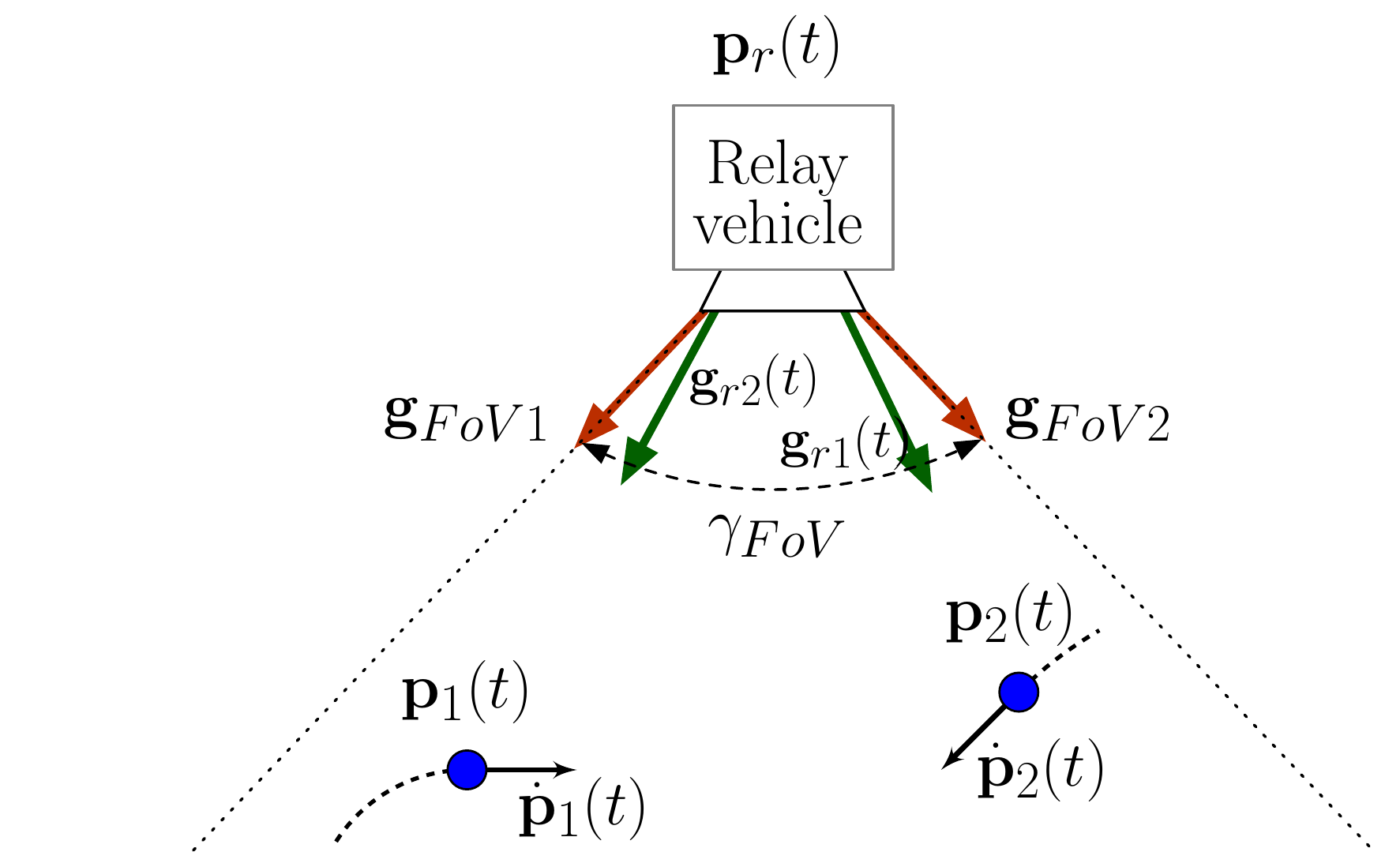}
	\caption{Top view illustration of the control problem.}
	\label{fig:relay}
\end{figure}
We also assume that the RV is equipped with one or more bearing measurement sensors with no range constraint but some FoV constraints.
The line-of-sight sensor for the RV is always facing towards a fix direction while the drone itself can move freely, as shown in Fig. \ref{fig:relay}.  %In this illustration, we see the position and velocity vectors of the two independent UAVs.  We denote their arbitrary trajectory by the black lines. 
The RV can obtain a bearing measurement to each drone (green arrows) taken w.r.t. a common reference frame $xy$.
The FoV constraint for the RV is characterized by two unit-length vectors $\mathbf{g}_{\scaleto{FoV1}{4pt}},\mathbf{g}_{\scaleto{FoV2}{4pt}} \in \mathbb{R}^{d}$ describing an angle $\gamma_{FoV}=2\gamma$, $\gamma \in (0,\pi/2]$, such that $$\mathbf{g}^{\top}_{\scaleto{FoV1}{4pt}}\mathbf{g}_{\scaleto{FoV2}{4pt}} = \cos_{2\gamma}, $$ with $\gamma$ denoting the angle between the bisector
\begin{equation}\label{eq:bisector}
	\g^{*} = \dfrac{\mathbf{g}_{\scaleto{FoV1}{4pt}}+\mathbf{g}_{\scaleto{FoV2}{4pt}}}{\left\|\mathbf{g}_{\scaleto{FoV1}{4pt}}+\mathbf{g}_{\scaleto{FoV2}{4pt}}\right\|},
\end{equation}
and one of the FoV vectors $\mathbf{g}_{\scaleto{FoV1}{4pt}}$, $\mathbf{g}_{\scaleto{FoV2}{4pt}}$.
Also, we define the distance between the RV and $i$-th agent as $d_{ri}(t) = \left\|\p_{i}(t)-\p_{r}(t)\right\|$ and, based on \eqref{eq:bearingvectordef}, we let $\mathbf{g}_{ri}(t) = (\p_{i}(t)-\p_{r}(t))/d_{ri}(t) \in \mathbb{R}^{d}$ be the unit vector pointing from the RV to the $i$-th agent, with $\g_{r}(t) = \begin{bmatrix}
	\g_{r1}^{\top}(t) & \cdots & \g_{rn}^{\top}(t)
\end{bmatrix}^{\top} \in \Rset{dn}$ denoting the ensemble vector containing all these relative bearing measurements. We say that the measurement $\mathbf{g}_{ri} $ falls inside the FoV domain if 
\begin{align*} \g_{ri} \in \Cmc = \{\g \in \mathbb{R}^d \, | \,&(\mathbf{R}_z(\pi/2)\,\mathbf{g}_{\scaleto{FoV1}{4pt}})\T \g \geq 0 \mbox{ and } \nonumber\\
	&(\mathbf{R}_z(\pi/2)\,\mathbf{g}_{\scaleto{FoV2}{4pt}})\T \g  \leq 0\},
\end{align*}
where
\begin{equation*}
	\mathbf{R}_z(\alpha) = \begin{bmatrix}
		\cos_\alpha & -\sin_\alpha  \\ \sin_\alpha & \cos_\alpha %& 0 \\ 0 & 0 & 1
	\end{bmatrix} ,
\end{equation*}
is the well-known rotation matrix expressing a vector rotation about the $z$ axis, forming a three-dimensional reference frame along with plane $xy$, by an angle $\alpha \in [0,2\pi)$. 

After these premises, the control problem we attempt to solve is then summarized as follows.
\begin{problem}\label{problem_statement}
	Design a control law $\u_{r}(t) = \u_{r}(\g_{ri}(t))$ based only on the bearing measurements $\g_{ri}(t)$, $i \in \Vmc$, such that it ensures all agents remain in the RV's FoV, that is $\g_{ri}(t) \in \mathcal{C}$, $\forall i\in \Vmc$, $\forall t > 0$ whenever $g_{ri}(0)\in\mathcal{C}$.
\end{problem}

In the next subsection we propose our control strategy; however, the following two hypotheses are also assumed henceforth in order to preserve the meaning of such bearing measurements over time and guarantee the tracking to begin from the initial time instant, respectively.

\begin{assumption}\label{asm:tracking_at_t0}
	At $t=0$, it holds that $\g_{ri}(0) \in \Cmc$, $\forall i\in \Vmc$.
\end{assumption}

\begin{assumption}\label{asm:coll_avoid_relay}
	There exists $\epsilon > 0$ such that $\forall t \geq 0$ it holds that $d_{ri}(t)  \geq \epsilon$, $\forall i\in \Vmc$.
\end{assumption}

In fact, Asm. \ref{asm:coll_avoid_relay} is not trivially satisfied.  In the following section we first develop control strategies assuming this assumption holds.  Then, in Sec. \ref{sec:collavoidimpl}, we augment our control strategies with a collision avoidance term that guarantees Asm. \ref{asm:coll_avoid_relay} holds.

\begin{remark}
		In practice, it takes a small time $t_{r} > 0$ for the RV to detect a moving target and calculate the corresponding guidance strategy. Consequently, if one agent is moving outside the FoV, it may escape the region before the RV has computed the corresponding control action. To this purpose, it is possible to define a transient region $\lambda \in (0,\gamma)$ nearby the boundary of the FoV cone for preventing such a scenario. In particular, given the highest admissible delay $T_{r} \geq t_{r}$, an estimate $\lambda^{*}$ for the smallest admissible transient region can be derived by imposing $ T_{r} v_{M} \leq \sin_{\lambda} \epsilon$. The term $\sin_{\lambda} \epsilon$ represents a lower bound estimate for the minimum escaping distance. For this reason, one obtains $\lambda^{*} = \mathrm{arcsin}(T_{r}v_{M}/\epsilon)$.
		A safer version $\gamma_{S} \in (0,\pi/2]$ of the FoV angle $\gamma$ can be  calculated as $\gamma_{S} = \gamma - \lambda$, with $\lambda \in [\lambda^{*}, \gamma)$, if $\epsilon$ is selected so that $\epsilon \geq T_{r} v_{M}$. The latter inequality needs to hold, since $T_{r} v_{M} / \epsilon \leq 1 $ is required for $\lambda^{*}$ to be well-defined.
		%\todo[inline]{above sentence is too long! can not parse it}
		Finally, the use of the angle $\gamma_{S}$ can replace\footnote{Throughout this paper, we consider to take the limit of $t_{r}$ towards $0$ neglecting the communication delay between RV and agents. Hence, for sake of simplicity, $T_{r}=0$ is set (leading to $\lambda^{*}=0$) and $\gamma_{S}=\gamma $ is adopted.} the adoption of angle $\gamma$ to ensure a suitable transient region. 
		%
		%Clearly, the latter observations also determine a couple of very intuitive trade-offs. One is that emerging between avoidance capability and communication delay: low values of $\epsilon$ are challenging to be adopted, while robustness of the system to high values of $T_{r}$ is desirable. On the other hand, there exists a trade-off between avoidance capability and maximum speed of the agents: collision avoidance for small values of $\epsilon$ and, simultaneously, large values of $v_{M}$ is hard to be guaranteed.
\end{remark}

\subsection{Control strategy} \label{ssec:control_strategy}
The general control strategy we propose is based on the bearing-only control law introduced in \eqref{bearing_control_law}.  In particular, we assume the RV controls its position based on the measurements of the other agents, and devise such bearing-based control law in three steps by taking into account the following tracking scenarios: a single agent, two agents and the general case of $n\geq 1$ agents. The details of these three cases are discussed in the sequel.

\subsubsection{The Single Agent Case}

In the single agent case ($n=1$), the control law for the RV takes the form
\begin{equation}\label{eq:ctrl_law_1}
	\u_{r}(\g_{r1}) = -K_{r} \P_{\g_{r1}} \g^{*},
\end{equation}
where $K_{r}>0$ is a control gain, $\P_{\g_{r1}}$ is the orthonormal projection operator defined in \eqref{eq:def_proj_matr_bearing}, evaluated at $\g_{r1}$, and $\g^{*}$ is the bisector characterized in \eqref{eq:bisector}. 
The fact that input \eqref{eq:ctrl_law_1} allows the RV to keep track of the sole agent $1$ is shown in the following proposition.

\begin{proposition}\label{prop:oneagent}
	Under Asm. \ref{asm:max_speed}, Asm. \ref{asm:tracking_at_t0}, Asm. \ref{asm:coll_avoid_relay} and the presence of a single agent, the adoption of control law \eqref{eq:ctrl_law_1} with $K_{r} \geq K_{r}^{\star} = v_{M}/\sin_{\gamma}$ implies that $\g_{r1}(t) \in \Cmc$, $\forall t\geq 0$. 
\end{proposition} 

\begin{figure}[h!]
	\centering
	\includegraphics[scale=0.2]{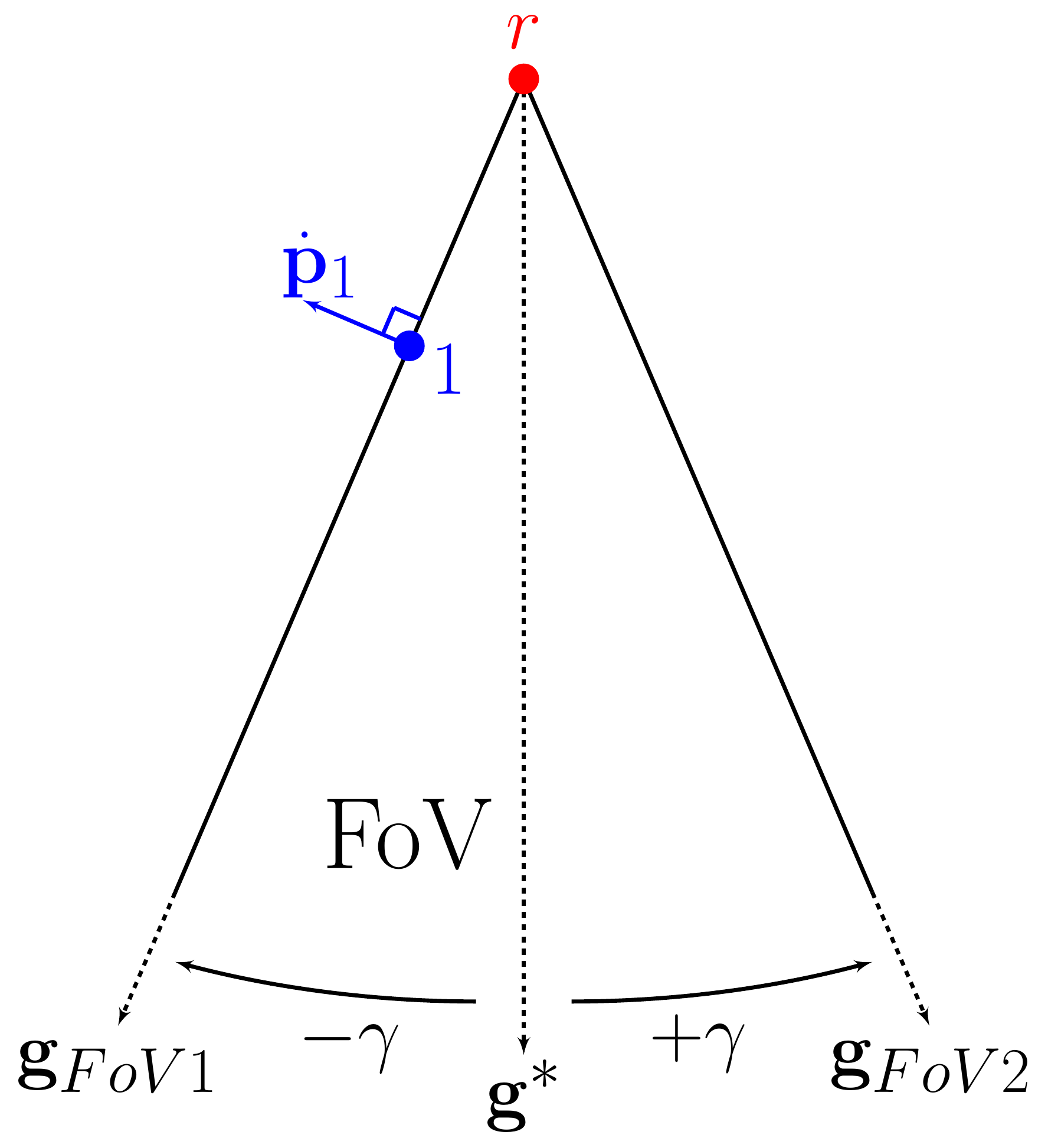}
	\caption{Single agent case scenario: geometric construction employed in Prop. \ref{prop:oneagent}. }
	\label{fig:singleagentproof}
\end{figure}

\begin{proof}
	It is sufficient to think of the only possible worst case scenario (see Fig. \ref{fig:singleagentproof}), in which agent $1$ is located along the FoV vector $\mathbf{g}_{\scaleto{FoV1}{4pt}}$, w.l.o.g., that is $\g_{r1} = \mathbf{g}_{\scaleto{FoV1}{4pt}}$, and is escaping with a velocity $\dot{\p}_{1}$ from the FoV, such that $\left\|\dot{\p}_{1}\right\|=v_{M}$, $\dot{\p}_{1}^{\top}\mathbf{g}_{\scaleto{FoV1}{4pt}}=0$ and $\dot{\p}_{1}^{\top}\mathbf{g}_{\scaleto{FoV2}{4pt}}\leq 0$.
	By leveraging the properties of projection operator $\P_{\g_{ri}}$ (see \eqref{eq:def_proj_matr_bearing}), one has $\mathbf{g}_{\scaleto{FoV1}{4pt}}^{\top} \P_{\g_{r1}} \g^{*} = \g_{r1}^{\top} \P_{\g_{r1}} \g^{*} = 0$; thus, vectors $\dot{\p}_{1}$ and $\P_{\g_{r1}} \g^{*}$ are parallel. %\todo[inline]{please just write it out mathematically rather than say "it is trivial"}
	Therefore, one can impose inequality
	$\left\|\u_{r}\right\| = K_{r} \left\| \P_{\g_{r1}} \g^{*} \right\| = K_{r} \sin_{\gamma} \geq v_{M} = \left\|\dot{\p}_{1}\right\|$ to compute the maximum speed effort needed by the RV to track agent $1$ and the thesis follows.
	%\todo[inline]{please generate a figure for this case - \MFa{Marco: what do you think about the content of this figure? Does it support properly the proof?}}
\end{proof}

\begin{figure}[t]
	\centering
	\includegraphics[scale=0.2]{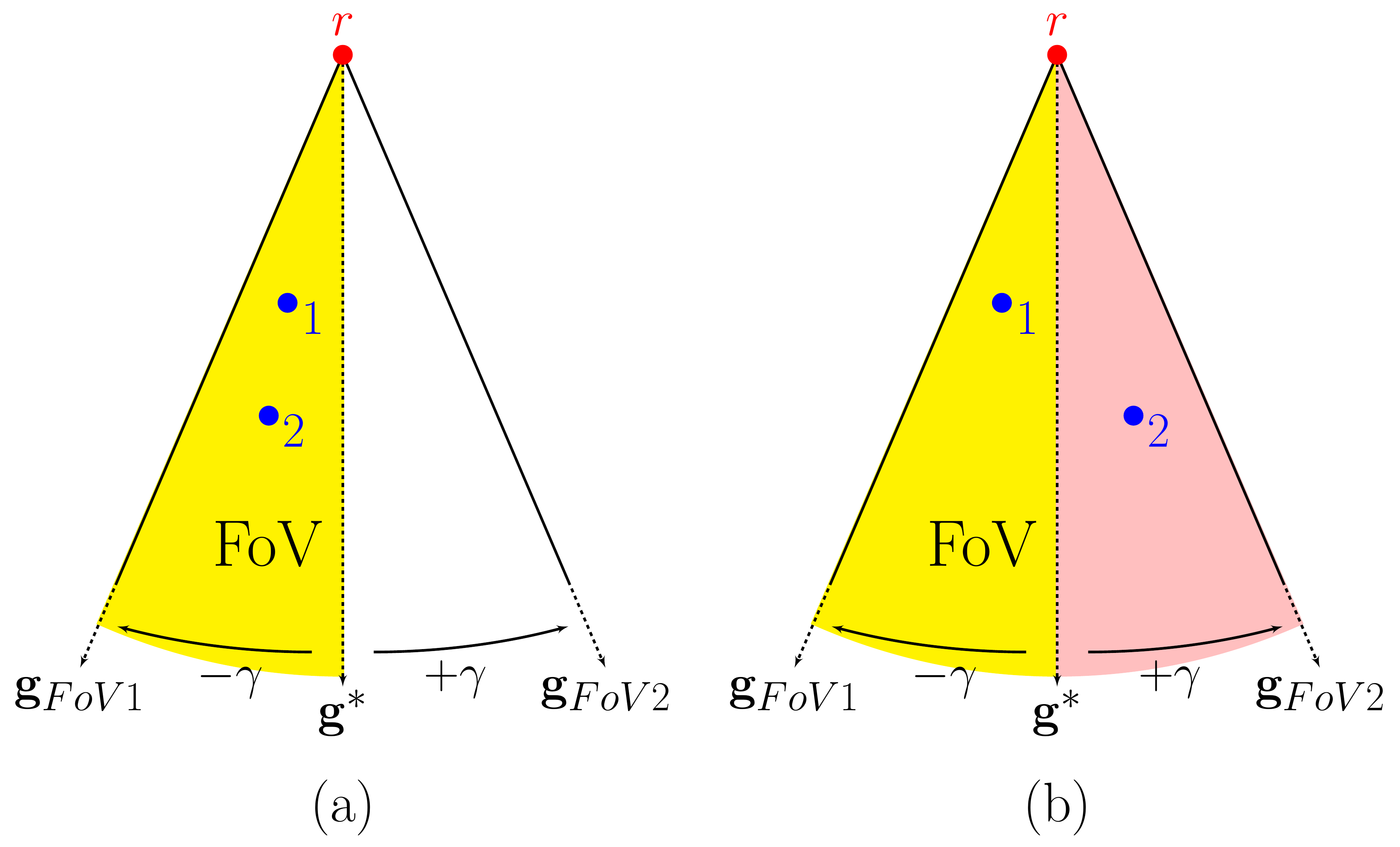}
	\caption{(a): Two agents belonging to the same side of the FoV (yellow). (b): Two agents laying on the different sides of the FoV (yellow and pink). } 
	\label{fig:twoagentsproof}
\end{figure}

$~$\\
\subsubsection{The 2-agent Case} 

Starting from the two agent case ($n=2$), there are two configurations of interest that specify the agent position relative to the RV.  As the relay FoV describes a cone, we consider the agent positions relative to the cone bisector, defined by vector $\g^{*}$ in \eqref{eq:bisector}. As illustrated in Fig. \ref{fig:twoagentsproof}, such positions can be distinguished by the \textit{side discriminator function} $\chi: \mathbb{R}^{2d} \to \{0, \pm 1\} $, such that
\begin{equation}\label{eq:sidediscriminator2}
	\chi(\g_{r}) =  \mathrm{sign}( \g_{r1}^{\top}\P_{\g^{*}}\g_{r2} ).
\end{equation}
Function in \eqref{eq:sidediscriminator2} returns $1$, if both the agents lay on the ``same side'', namely they belong to the angular portion delimited by the bisector $\g^{*}$ and one of the bearing vectors $\mathbf{g}_{\scaleto{FoV1}{4pt}}, \mathbf{g}_{\scaleto{FoV2}{4pt}}$. It returns $-1$, if the two agents lay on ``opposite sides'', namely one of them belongs to the angular portion between $\g^{*}$ and $\mathbf{g}_{\scaleto{FoV1}{4pt}}$ and the other belongs to the angular portion between $\g^{*}$ and $\mathbf{g}_{\scaleto{FoV2}{4pt}}$. It returns $0$, whenever there exist at least one agent laying on the bisector $\g^{*}$. See also the next paragraph for a more precise, formal and comprehensive description of the side discriminator function in a scenario presenting a generic number of agents.

In this direction, we propose a switching controller for the RV.  The main idea is to follow the control strategy given in \eqref{eq:ctrl_law_1}, when both agents are on the same side of the bisector $\g^{*}$, and switch to a controller that utilizes both measurements, when the agents are on opposite sides of $\g^{*}$.  
If $\chi \geq 0$, we denote with $\bar{\g}_{r} \in \Rset{d}$ the unit vector that points toward the agent closest to $\mathbf{g}_{\scaleto{FoVj}{4pt}}$, depending on the side $j$ where agents lay, namely
\begin{align}\label{eq:close_bearing}
	\bar{\g}_r =  \underset{\{\g_{r1},\g_{r2}\}}{\arg \max}  \left\lbrace   \underset{j\in \{1,2\}}{\max} \g_{ri}^{\top}\mathbf{g}_{\scaleto{FoVj}{4pt}} \right\rbrace. %\arg  \max_i \mathbf{g}_{\scaleto{FoVj}{4pt}}^T \g_{ri}, \, j=1,2.
\end{align}

We are now prepared to present the switching controller for the scenario with $n=2$ agents:
\begin{align}\label{eq:ctrl_law_2}
	\u_{r}(\g_{r})  & =\begin{cases}
		-K_{r}\P_{\bar{\g}_r} \g^{*}, & \text{ if }\chi(\g_{r}) \geq 0;\\
		-K_{r}(\P_{\g_{r1}}+\P_{\g_{r2}}) \g^{*},& \text{ if }\chi(\g_{r}) < 0.
	\end{cases}  
\end{align}
The validity of this control law, i.e., the fact that \eqref{eq:ctrl_law_2} allows the RV to keep track of all agents in the FoV for all $t \geq 0$, is proven in the following lemma.

\begin{lemma}\label{lemma:twoagents}	
	Let us define the real positive quantity
		\begin{equation}\label{eq:qstarvalue}
			q_{\gamma}^{*} = \begin{cases}
				2 \sin_{\gamma}^{3}, &\quad \text{if } \gamma \in \left(  0,\frac{\pi}{6} \right]; \\
				\frac{3}{2}\sin_{\gamma}-\frac{1}{2}, &\quad \text{if } \gamma \in \left( \frac{\pi}{6}, \frac{\pi}{2} \right].
			\end{cases}
	\end{equation}
	Then, under Asm. \ref{asm:max_speed}, Asm. \ref{asm:tracking_at_t0}, Asm. \ref{asm:coll_avoid_relay} and the presence of two agents, the adoption of control law \eqref{eq:ctrl_law_2} with $K_{r} \geq  v_{M}/q_{\gamma}^{*}$ implies that $\g_{ri}(t) \in \Cmc$, for $ i = 1,2$, $\forall t \geq 0$.
\end{lemma}

\begin{proof}
	The proof can be split into a couple of macro cases, as control law \eqref{eq:ctrl_law_2} switches according to $\chi$.
	
	If $\chi \geq 0$ then Prop. \ref{prop:oneagent} applies by considering only the agent $\bar{k} \in \{1,2\}$ corresponding to bearing $\bar{\g}_{r}$. Indeed, this is the closest agent -- in terms of angles -- to one of the FoV vectors $\mathbf{g}_{\scaleto{FoV1}{4pt}}$, $\mathbf{g}_{\scaleto{FoV2}{4pt}}$, since maximization in \eqref{eq:close_bearing} is equivalent to the minimization of the angle between $\g_{ri}$ and $\mathbf{g}_{\scaleto{FoVj}{4pt}}$, defined as
	\begin{align*}\label{eq:phi}
		\phi_i^{(j)} = \arccos(\g_{ri}^{\top}\mathbf{g}_{\scaleto{FoVj}{4pt}}) , \quad j=1,2.
	\end{align*} 
	Consequently, the trajectory of the other agent (the one different from $\bar{k}$) can be neglected by controller $\u_{r}$, as $\chi \geq 0$ and, therefore, the thesis follows.
	
	\begin{figure}[!t]
		\centering
		\includegraphics[scale=0.215]{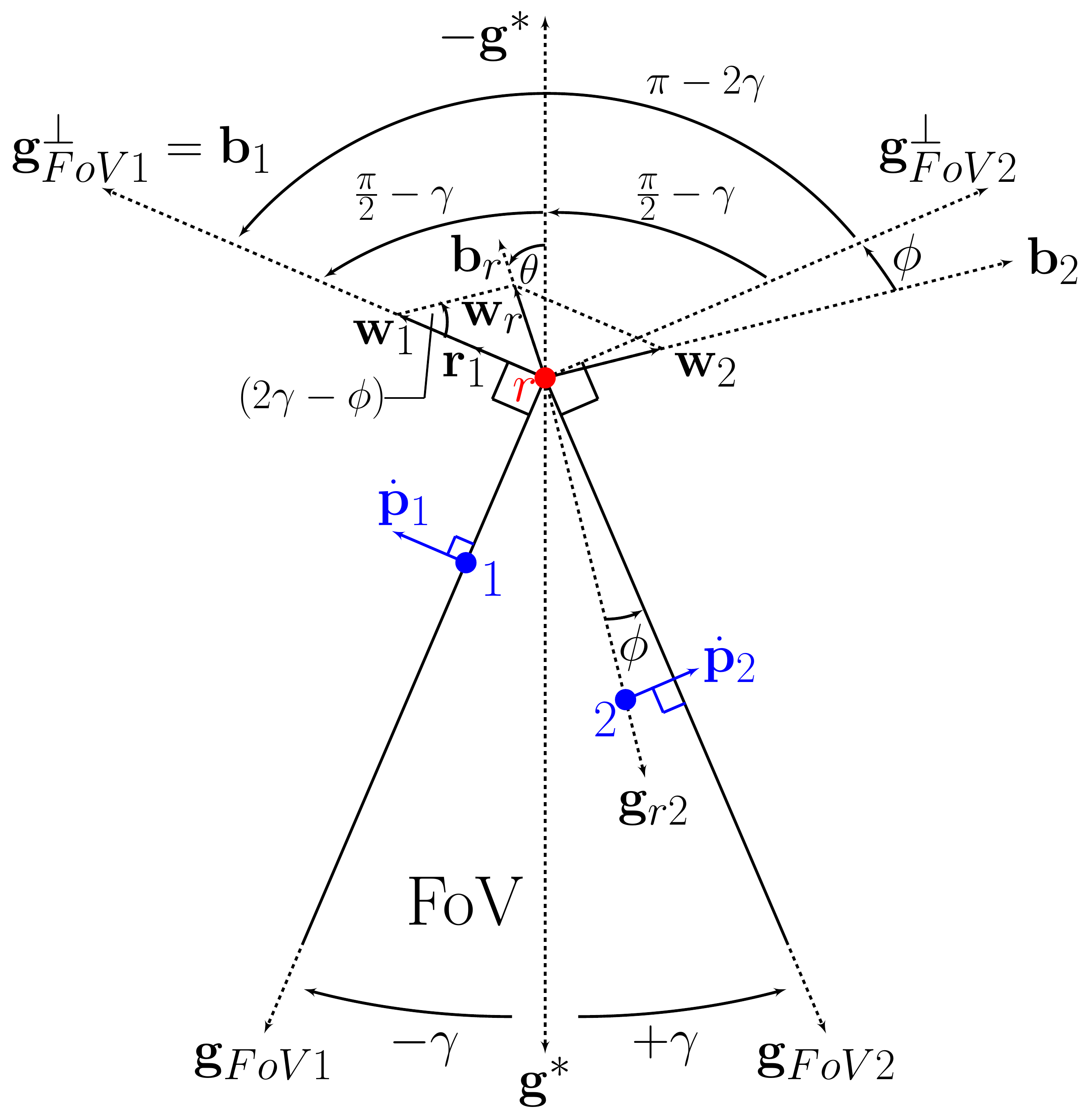}
		\caption{Two-agent case scenario: geometric construction employed in Lem. \ref{lemma:twoagents}. 
			Unit vectors $\mathbf{g}_{\scaleto{FoVj}{4pt}}^{\perp} = \R_{z}((-1)^{j}\pi/2) \mathbf{g}_{\scaleto{FoVj}{4pt}}$, with $j=1,2$, are also defined to help the visualization.}
		\label{fig:max_velocity}
	\end{figure}
	
	Otherwise, if $\chi < 0$ then we think of the worst case scenario in which agent $1$ is located, again, w.l.o.g., along the FoV vector $\mathbf{g}_{\scaleto{FoV1}{4pt}}$ and is escaping with a velocity $\dot{\p}_{1}$ such that $\left\|\dot{\p}_{1}\right\| = v_{M}$, $\dot{\p}_{1}^{\top}\mathbf{g}_{\scaleto{FoV1}{4pt}}=0$ and $\dot{\p}_{1}^{\top}\mathbf{g}_{\scaleto{FoV2}{4pt}}\leq 0$.\\
	Now, the proof requires the geometric constructions depicted in Fig. \ref{fig:max_velocity}. Let us define $\w_{1} = -\P_{\g_{r1}} \g^{*} = \left\|\w_{1}\right\|\b_{1}$ and $\w_{2} = -\P_{\g_{r2}} \g^{*} = \left\|\w_{2}\right\| \b_{2}$, such that $\left\|\b_{1}\right\| = \left\|\b_{2}\right\| = 1$. By the properties of projection operators $\P_{\g_{ri}}$ it holds that $\left\|\w_{1}\right\|=\sin_{\gamma}$, $\left\|\w_{2}\right\|=\sin_{\gamma-\phi}$, assigning $\phi = \min_{j\in \{1,2\}} \phi_{2}^{(j)}$. The vector addition $\w_{r}$ between $\w_{1}$ and $\w_{2}$ is thus given by $\w_{r}=\sin_{\gamma}\b_{1} + \sin_{\gamma-\phi} \b_{2}$. Since the angle (between $0$ and $\pi$) taken from $\b_{1}$ to $\b_{2}$ is equal to $(\pi-2\gamma+\phi)$, then one has 
	\begin{equation}\label{eq:wrnorm}
			\left\|\w_{r}\right\|^{2} = \sin_{\gamma}^{2}+\sin_{\gamma-\phi}^{2}-2\cos_{2\gamma-\phi}\sin_{\gamma}\sin_{\gamma-\phi}.
	\end{equation}
	Therefore, redefining $\w_{r} = \left\|\w_{r}\right\| \b_{r}$, with $\left\|\b_{r}\right\|=1$, it is possible to rewrite case $\chi <0$ in \eqref{eq:ctrl_law_2} alternatively as $\dot{\p}_{r} = K_{r} \left\|\w_{r}\right\| \b_{r}$.\\
	At the light of these observations, we distinguish between two nested cases. Case (i) deals with agent $1$ solely escaping from the FoV, while case (ii) copes with both agents $1$ and $2$ exiting the FoV.
	
	Case (i): $\phi \in (0,\gamma]$. According to this configuration, the RV is required to track agent $1$ only. Hence, we consider the projection $\r_{1}$ of RV's velocity $\dot{\p}_{r}$ onto the direction of $\dot{\p}_{1}$ and impose that the minimum among all potential speeds $\left\|\r_{1}\right\|$ be greater than $v_{M}$, as $\phi$ varies. 
	Denoting with $\theta=\theta(\gamma,\phi)\in [0,\pi/2-\gamma]$ the angle between $-\g^{*}$ and $\b_{r}$, such that $-\b_{r}^{\top}\g^{*} = \cos_{\theta}$, the norm of $\r_{1}$ is yielded by
	\begin{equation}\label{eq:normofr1}
			\left\|\r_{1}\right\|= K_{r} \left\|\w_{r}\right\| \cos_{\pi/2-\gamma-\theta} = K_{r}   \sin_{\gamma}  \cos_{\theta}\left\|\w_{r}\right\| +K_{r} \cos_{\gamma} \sin_{\theta}\left\|\w_{r}\right\| .
		\end{equation}
		Moreover, from the fact that
		%	\begin{align*}
			%		\cos_{\theta}\left\|\w_{r}\right\| &= -\w_{r}^{\top}\g^{*} \\
			%		&= \sin_{\gamma} \b_{1}^{\top} (-\g^{*}) +\sin_{\gamma-\phi} \b_{2}^{\top} (-\g^{*}) \\
			%		&= \sin_{\gamma}\cos_{\pi/2-\gamma} + \sin_{\gamma-\phi}\cos_{\phi+\pi/2-\gamma} = \sin_{\gamma}^{2}+\sin_{\gamma-\phi}^{2},
			%	\end{align*}
		\begin{equation*}
			\cos_{\theta}\left\|\w_{r}\right\| = -\w_{r}^{\top}\g^{*} = \sin_{\gamma} \b_{1}^{\top} (-\g^{*}) +\sin_{\gamma-\phi} \b_{2}^{\top} (-\g^{*}) = \sin_{\gamma}\cos_{\pi/2-\gamma} + \sin_{\gamma-\phi}\cos_{\phi+\pi/2-\gamma} = \sin_{\gamma}^{2}+\sin_{\gamma-\phi}^{2},
		\end{equation*}
		the expressions $\cos_{\theta}\left\|\w_{r}\right\|=\sin_{\gamma}^{2}+\sin_{\gamma-\phi}^{2}$ and $\sin_{\theta}\left\|\w_{r}\right\| = \sqrt{ \left\|\w_{r}\right\|^{2}-\cos_{\theta}^{2}\left\|\w_{r}\right\|^{2}}$ can be written explicitly as a  function of $\gamma$ and $\phi$ only. Indeed, substituting such expressions and \eqref{eq:wrnorm} in \eqref{eq:normofr1}, one obtains $\left\|\r_{1}\right\| = K_{r} q_{\gamma}(\phi)$,
		where function $q_{\gamma}:[0,\gamma] \rightarrow [q_{\gamma}^{*},\max(2\sin_{\gamma}^{3},\sin_{\gamma})] \subset (0,2]$ is defined as
		\begin{equation}\label{eq:qgammaofphi}
			q_{\gamma}(\phi) =  
			\sin_{\gamma}^{3} +\sin_{\gamma}\sin_{\gamma-\phi}^{2} + \cos_{\gamma} \sqrt{\sin_{\gamma}^{2}+\sin_{\gamma-\phi}^{2}-2\cos_{2\gamma-\phi}\sin_{\gamma}\sin_{\gamma-\phi}-(\sin_{\gamma}^{2}+\sin_{\gamma-\phi}^{2})^{2}}.
		\end{equation}
		As demonstrated in App. \ref{sec:appendix}, quantity $q_{\gamma}^{*}$ actually coincides with the minimum of \eqref{eq:qgammaofphi}. Also, it is worth to notice that $q_{\gamma}^{*} \leq \sin_{\gamma}$ for all values of $\gamma \in (0,\pi/2]$.
	Hence, the control gain selection can be done by setting $\min_{\phi\in (0,\gamma]}\left\|\r_{1}\right\|= K_{r} q_{\gamma}^{*} \geq v_{M}$.

	Case (ii): $\phi = 0$. This configuration is slightly different because the position of agent $2$ is exactly located on the $\mathbf{g}_{\scaleto{FoV2}{4pt}}$ direction. Hence, the RV may lose tracking of agent $1$, agent $2$ or both of them. For this reason, we add a further consideration to the previous premises in order to tackle the worst case scenario arising when $\phi = 0$: we assume that agent $2$ is also escaping with a velocity $\dot{\p}_{2}$ from the FoV, so that $\left\|\dot{\p}_{2}\right\|=v_{M}$, $\dot{\p}_{2}^{\top}\mathbf{g}_{\scaleto{FoV2}{4pt}}=0$ and $\dot{\p}_{2}^{\top}\mathbf{g}_{\scaleto{FoV1}{4pt}}\leq 0$. 
	
	Besides this preliminary observation, the geometric constructions discussed for the previous case $\phi \in (0,\gamma]$ remain valid here. In particular, one has $\left\|\r_{1}\right\| = K_{r} q_{\gamma}(0) = 2K_{r}\sin_{\gamma}^{3}$. Similarly, defining $\r_{2}$ as the projection of $\w_{r}$ onto the direction of $\dot{\p}_{2}$, it holds that $\left\|\r_{2}\right\| = \left\|\r_{1}\right\|$, by geometrical symmetry ($\w_{r}$ lays on the same direction of the bisector $-\g^{*}$). Therefore, inequality $2K_{r}\sin_{\gamma}^{3} \geq v_{M}$ leads to a particular control gain selection for $\phi=0$. However, the latter inequality cannot be used as a general condition for the control gain selection because for $\gamma \in (\pi/6,\pi/2]$ it holds that $q_{\gamma}^{*} = (3\sin_{\gamma}-1)/2 < 2\sin_{\gamma}^{3} = q_{\gamma}(0)$. Thus, imposing again $K_{r}q_{\gamma}^{*} \geq v_{M}$ concludes the proof. 
\end{proof}

\subsubsection{Generalization for $n>2$ agents}
We finally discuss our main contribution, a bearing-based control law for the autonomous tracking of $n>2$ agents. To provide such a formalization, we need to introduce a few new quantities. Firstly, we determine whether an agent $i\in \Vmc$ is moving in the left or right part of the FoV w.r.t. bisector $\g^{*}$ defined in \eqref{eq:bisector}. Alternatively, $i$ may also be located exactly at the FoV bisector. To do so, we first define set $\mathbb{M}=\{1,2,3\}$ and function $s_{i}:\Rset{d} \rightarrow \mathbb{M}$, such that
\begin{equation}\label{eq:sigmai}
	s_{i}(\g_{ri}) = \begin{cases}
		1, \quad \g_{ri}^{\top}(\mathbf{g}_{\scaleto{FoV2}{4pt}}-\mathbf{g}_{\scaleto{FoV1}{4pt}}) = 0;\\
		2, \quad \g_{ri}^{\top}(\mathbf{g}_{\scaleto{FoV2}{4pt}}-\mathbf{g}_{\scaleto{FoV1}{4pt}}) < 0;\\
		3, \quad \g_{ri}^{\top}(\mathbf{g}_{\scaleto{FoV2}{4pt}}-\mathbf{g}_{\scaleto{FoV1}{4pt}}) > 0;\\
	\end{cases}
\end{equation}
which returns $s_{i}=2$, if $\p_i$ belongs to the left half the FoV domain (planar portion between $\mathbf{g}_{\scaleto{FoV1}{4pt}}$ and $\g^{*}$, i.e. the left side); otherwise, $s_{i}=3$, if $\p_i$ belongs to the right half the FoV domain (planar portion between $\mathbf{g}_{\scaleto{FoV2}{4pt}}$ and $\g^{*}$, i.e., the right side). Alternatively, $s_{i}=1$ is taken, if $\p_i$ belongs to the bisector along $\g^{*}$.\\
Subsequently, setting $\s = \s(\g_{r}) = \begin{bmatrix}
	s_{1}(\g_{r1}) & \cdots & s_{n}(\g_{rn})
\end{bmatrix}^{\top} \in \mathbb{M}^{n}$, we also define functions $\sigma_{1},\sigma_{2},\sigma_{3} : \mathbb{M}^{n} \to \mathbb{M}$, such that
\begin{equation}
	\sigma_{j}(\s) = |\{i\in \mathcal{V} \, | \, s_{i}=j\}|, \quad j=1,2,3,
\end{equation}
yielding respectively for $j=1,2,3$ the number of agent laying on the bisector, left side and right side.

Then, setting $\Bsigma = \Bsigma(\s) = \begin{bmatrix}
	\sigma_{1}(\s) & \sigma_{2}(\s) & \sigma_{3}(\s)
\end{bmatrix}^{\top} \in \mathbb{M}^{3}$, we are now allowed to generalize the side discriminator function in \eqref{eq:sidediscriminator2} for $n$ agents via function $ \xi_{n} : \mathbb{M}^{3} \to \{0, \pm 1\}$ having the form
\begin{equation}\label{eq:xi}
	\xi_{n}(\Bsigma) = \begin{cases}
		1, \quad \text{ if } 2 \leq \max \{\sigma_{2},\sigma_{3}\} = n-\sigma_{1} \leq n; \\
		0, \quad \text{ if } n-1 \leq \sigma_{1} \leq n; \\
		-1, \quad \!\!\!\! \text{ otherwise};
	\end{cases}
\end{equation}
by finally assigning the function composition that extends $\chi(\g_{r})$ into $\chi_{n}(\g_{r}) := \xi_{n}(\Bsigma(\s(\g_{r})))$.
Indeed, in its characterization, one configuration consists in at least two agents laying on the ``same side'' of the vector $\g^{*}$ (for $\chi_{n} = 1$), with all the others, if any, laying exactly on the bisector. This means, more formally, that the maximum length of projection $\g_{ri}^{\top} \mathbf{g}_{\scaleto{FoVj}{4pt}}$ is attained either for $j=1$ or $j=2$, for each agent $i \in \bar{\Vmc} \subset \Vmc$, with $\bar{\Vmc} = \{i\in \Vmc \; | \; \g_{ri} \neq \g^{*}\}$, $|\bar{\Vmc}| \geq 2$.
Whereas, another configuration ($\chi_{n} = -1$) describes agents located on ``opposite sides'' of the bisector, i.e. there exist $\bar{k}_{1},\bar{k}_{2} \in \Vmc$, with $\bar{k}_{1}\neq \bar{k}_{2}$, such that $\g_{ri} \neq \g^{*}$ for $i = \bar{k}_{1}, \bar{k}_{2}$, and the length of projection $\g_{ri}^{\top} \mathbf{g}_{\scaleto{FoVj}{4pt}}$ is maximized for $j=1$, if $i=\bar{k}_{1}$, and for $j=2$, if $i=\bar{k}_{2}$. Alternatively, $\chi_{n}$ takes $0$ value whenever these two possibilities lose proper meaning.
In particular, the following result holds.
\begin{proposition}\label{prop:chiequiv}
	For the case $n=1$, one has $\chi_{1}(\g_{r}) = 0$, $\forall \g_{r}\in \Rset{d}$.
	Also, for the case $n=2$, one has $\chi_{2}(\g_{r}) = \chi(\g_{r})$, $\forall \g_{r}\in \Rset{d}$. 
\end{proposition}

\begin{proof}
	The first part of the statement is trival, since, if $n=1$, then condition $n-1 \leq \sigma_{1} \leq n$ in \eqref{eq:xi} is ensured to hold true because either $ \sigma_{1} =0$ or $ \sigma_{1} = 1$.
	The second part of the statement can be proven by observing the fact that $s_{i} $ in \eqref{eq:sigmai} is equal to $1$ if and only if $\g_{ri}$ and $\g^{*}$ are parallel, namely $\g_{ri} = \g^{*}$. Then, assuming $n=2$ and focusing on the claim $\chi_{2} = \chi = 0$, condition $n-1 \leq \sigma_{1} \leq n$ in \eqref{eq:xi} holds if and only if $\g_{r1} = \g^{*}$ or $\g_{r2} = \g^{*}$, i.e. if and only if term $\g_{r1}^{\top}\P_{\g^{*}}\g_{r2}$ in \eqref{eq:sidediscriminator2} is null. Moreover, given the previous conclusion on $\chi_{2} = \chi = 0$, it is possible to prove a correspondence between $\chi_{2}$ and $\chi$ when $\chi_{2} = \chi = 1$ is claimed. Indeed, if $\chi_{2} = 1$, then $\sigma_{1} = 0$ is forced, as $1 \leq \sigma_{1} \leq 2$ cannot be possible. Therefore, condition $2 \leq \max \{\sigma_{2},\sigma_{3}\} = n-\sigma_{1} \leq n$ in \eqref{eq:xi} boils down to $\max \{\sigma_{2},\sigma_{3}\}=2$. Since $n=2$, then either $\sigma_{2}=2$ or $\sigma_{3}=2$, meaning that both the agents lay on one of the sides ($\chi = 1$) and the thesis follows.
\end{proof}

The development of the general control law continues by computing $\bar{\g}_{r} \in \Rset{d}$ similarly to \eqref{eq:close_bearing}, namely through 
\begin{align}\label{eq:close_bearing_n}
	\bar{\g}_r =  \underset{\{\g_{r1},\ldots,\g_{rn}\}}{\arg \max}  \left\lbrace   \underset{j\in \{1,2\}}{\max} \g_{ri}^{\top}\mathbf{g}_{\scaleto{FoVj}{4pt}} \right\rbrace. 
\end{align}
and defining vector $\bar{\g}_{rj} \in \Rset{d}$ as
\begin{equation}\label{eq:grjbar}
	\bar{\g}_{rj} = \underset{\{\g_{r1},\ldots,\g_{rn}\}}{\arg \max}  \g_{ri}^{\top}\mathbf{g}_{\scaleto{FoVj}{4pt}}, \quad j=1,2.
\end{equation}
Unit vectors in \eqref{eq:grjbar} identify with both the closest bearings w.r.t. the two FoV vectors $\mathbf{g}_{\scaleto{FoV1}{4pt}}$, $\mathbf{g}_{\scaleto{FoV2}{4pt}}$.

With the geometrical entities introduced above, we are now ready to propose the general control law for the RV to maintain the agents inside its FoV:
\begin{align}\label{eq:bearing_general_control} 
	\u_{r}(\g_{r})  & =\begin{cases}
		-K_{r}\P_{\bar{\g}_r} \g^{*}, & \text{ if }\chi_{n}(\g_{r}) \geq 0;\\
		-K_{r} \left(\P_{\bar{\g}_{r1}}+\P_{\bar{\g}_{r2}}\right) \g^{*},& \text{ if }\chi_{n}(\g_{r}) < 0.
	\end{cases}  
\end{align}
The next theorem discusses this general case of autonomous relay tracking in details.
\begin{theorem}\label{thm:control_law_general}
	Let $q_{\gamma}^{*}$ be defined as in \eqref{eq:qstarvalue}. Under Asm. \ref{asm:max_speed}, Asm. \ref{asm:tracking_at_t0}, Asm. \ref{asm:coll_avoid_relay} and the presence of $n\geq 1$ agents, the adoption of control law \eqref{eq:sigmai}-\eqref{eq:bearing_general_control} with $K_{r} \geq K_{r}^{q} = v_{M}/q_{\gamma}^{*}$ solves Prob. \ref{problem_statement}, i.e. it implies that $\g_{ri}(t) \in \Cmc$, for $ i\in 1,\ldots,n$, $\forall t \geq 0$.
\end{theorem}

\begin{proof}
	The proof is again faced by splitting the analysis into two macro cases according to control law \eqref{eq:bearing_general_control} and leveraging Prop. \ref{prop:chiequiv}. 
	
	Case $\chi_{n} \geq 0$. Denoting with $\bar{k}$ the agent corresponding to $\bar{\g}_r$ and recalling case $\chi_{2} = \chi \geq 0$ in Lem. \ref{lemma:twoagents}, similar conclusions can be drawn trivially to show the thesis. Indeed, all the trajectories of the agents different from $\bar{k}$ can be neglected and therefore Prop. \ref{prop:oneagent} applies to $\bar{k}$. However, it is worth to notice that even though dynamics provided by input \eqref{eq:bearing_general_control} in this case boils down to \eqref{eq:ctrl_law_1}, the gain condition $K_{r}\geq v_{M}/q_{\gamma}^{*}$ is imposed. Such a selection for the control gain complies with the requirements of Prop. \eqref{prop:oneagent}, since it holds that $q_{\gamma}^{*} \leq \sin_{\gamma} $ for all $\gamma \in (0,\pi/2]$, yet it is more conservative in general (but also strictly necessary, as we will see for case $\chi_{n} < 0$). According to Prop. \ref{prop:minimizerq} in App. \ref{sec:appendix}, this fact can be proven by noting that $q_{\gamma}(\gamma)=\sin_{\gamma}$ and $q_{\gamma}^{\prime}(\gamma) = \cos_{\gamma} > 0$ for all $\gamma \in (0,\pi/2)$. Whereas, for $\gamma=\pi/2$, equality $q_{\gamma}(\gamma)=\sin_{\gamma}^{3}=1=\sin_{\gamma}$ implies that the control gain selection is exactly equivalent to that in Prop. \ref{prop:oneagent}. 
	%\todo[inline]{I don't understand what is meant by "according to what debated in Sec. 6...}
	
	Case $\chi_{n}<0$. Let us denote with $\bar{k}_{1}$ and $\bar{k}_{2}$ the agents associated to $\bar{\g}_{r1}$ and $\bar{\g}_{r2}$, respectively. Here, Lem. \ref{lemma:twoagents} can be applied neglecting all the agents different from $(\bar{k}_{1},\bar{k}_{2})$. Indeed, in this scenario, dynamics provided by input \eqref{eq:bearing_general_control} boils down to \eqref{eq:ctrl_law_2}, under case $\chi_{2} = \chi <0$, by treating $(\bar{k}_{1},\bar{k}_{2})$ as agents $1$ and $2$ in said lemma.
\end{proof}

Prop. \ref{prop:chiequiv} also implies that the formulation in \eqref{eq:bearing_general_control} is consistent with \eqref{eq:ctrl_law_1} in the single agent case and argument in Thm. \ref{thm:control_law_general} reduces exactly to Prop. \ref{prop:oneagent} and Lem. \ref{lemma:twoagents} as soon as $n=1$ and $n=2$, respectively, due to the lack of further agents. Nevertheless, Prop. \ref{prop:oneagent} remains a standalone theoretical result since it allows for a less conservative control gain selection in the single agent case. To conclude, a final remark is given.
\begin{remark}
	Control law provided in \eqref{eq:bearing_general_control} is not affected by discontinuities as $\chi_{n}$ changes, since $\P_{\g^{*}} \g^{*} = \zeros_{}$. This implies that undesired chattering phenomena related to sudden unexpected oscillations in the RV dynamics do not arise.
\end{remark}

\subsection{Collision avoidance implementation}\label{sec:collavoidimpl}

The control strategy devised in Sec. \ref{ssec:control_strategy} allows the RV to maintain all agents inside its FoV over time.  However, possible collisions among the RV and other vehicles have been neglected so far to simplify the setup and obtain the sharp theoretical results previously discussed. Therefore, in this section, we intend to develop a more suitable control strategy, able to guarantee the validity of Asm. \ref{asm:coll_avoid_relay} in practice. To this aim, short-range distance sensing is required to be embedded in the RV in order to implement some safety measure that mitigates the severity of potential crashes between robots. In this work we assume the RV is able to measure also distances to other agents that are with an $\epsilon_{s}$-ball of the RV.   Furthermore, we denote by $\epsilon \in (0, \epsilon_{s})$ the minimum safety distance to the RV that ensures no collisions.
In this direction, dynamics in \eqref{integrator_dynamics} is modified into
\begin{equation}\label{eq:dynca}
	\dot{\p}_{r}(t) = \u_{r}(t) + \Bupsilon_{r}(t),
\end{equation}
wherein control input $\Bupsilon_{r}(t) \in \mathbb{R}^{d}$ provides a collision avoidance term and $\u_{r}(t) = \u_{r}(\g_{r}(t))$ is taken as in \eqref{eq:ctrl_law_1} or \eqref{eq:bearing_general_control}. From a design point of view, we can already identify the main features of such collision avoidance term, i.e., defining 
\begin{equation}
	\Vmc_{s}(t) = \{i \in \Vmc ~|~ d_{ri}(t)\leq \epsilon_{s}\}
\end{equation}
and
\begin{equation}\label{eq:dr}
	d_{r}(t) = \begin{cases}
		\min_{i \in  \Vmc_{s}} \{d_{ri}(t)\}, & \text{if } \Vmc_{s}(t) \neq \emptyset; \\
		\epsilon_{s}, & \text{otherwise}.
	\end{cases}
\end{equation}
We can impose the factorization 
\begin{equation} \label{eq:collav_factors}
	\Bupsilon_{r}(t) = -\eta_{\epsilon}(d_{r}(t))a_{r}(t)\f_{r}(t), 
\end{equation}
in which $\eta_{\epsilon} : d_{r}(t) \in [0,+\infty) \mapsto \eta_{\epsilon}(d_{r}(t)) \in [0,1]$ is a \textit{collision alert function} whose support is a subset of $[0,\epsilon_{s}]$, $a_{r}(t) \geq 0$ is the \textit{avoidance effort} and $-\f_{r}(t)\in \mathbb{R}^{d}$ is a unit vector representing the \textit{escaping direction}. In particular, function $\eta_{\epsilon}(d_{r}(t))$ is the only factor in \eqref{eq:collav_factors} depending on short-range distance measurements and its design has a high degree of freedom. For sake of simplicity, we only consider collision alert functions that can be generally characterized as: 
\begin{align}\label{eq:etaeps_alertfun}
	\eta_{\epsilon}(d_{r}(t)) = \begin{cases}
		1, & 0\leq d_r(t) \leq \epsilon; \\
		\bar{\eta}_{\epsilon}(d_{r}(t)), & \epsilon < d_r(t) < \epsilon_s; \\
		0, & \text{otherwise}.
	\end{cases}
\end{align}
For $\epsilon < d_{r} < \epsilon_{s}$, function $\eta_{\epsilon}(d_{r})$ can be determined in different fashions to render continuous the transition towards the \textit{activation phase} $d_{r} \leq \epsilon$ and \textit{deactivation phase} $d_{r} \geq \epsilon_{s}$, e.g. by taking $\bar{\eta}_{\epsilon}(d_{r})$ nonincreasing in $d_{r}$ and such that $\lim_{d_{r} \rightarrow \epsilon^{+}} \bar{\eta}_{\epsilon}(d_{r}) = 1$ and $\lim_{d_{r} \rightarrow \epsilon_{s}^{-}} \bar{\eta}_{\epsilon}(d_{r}) = 0$. % with a fast decay rate as $d_{r}$ grows.

Some further preliminary definitions are needed for the discussion. Let us denote with 
\begin{equation}\label{eq:Vrset}
	\Vmc_{r}(t) = \{ i \in \Vmc_{s}(t) ~|~ d_{ri}(t)  = d_{r}(t) \}
\end{equation}
the set containing the sole agents to be addressed for accomplishing avoidance maneuvers. 
In particular, if $\Vmc_{r}(t) \neq \emptyset$, we account for agents 
\begin{equation*}
	(i^{*}(t),j^{*}(t)) = \underset{(i,j)\in \Vmc_{r}(t) \times \Vmc_{r}(t) \text{ and } i\leq j}{\arg \min}  \{ \g_{ri}(t)^{\top}\g_{rj}(t)\}
\end{equation*}
belonging to $\Vmc_{r}(t)$ in order to effectively build a collision avoidance strategy.
Note that $i^{*}(t)$ and $j^{*}(t)$ may coincide, so we define set $\Vmc_{r}^{*}(t) = \{i^{*}(t),j^{*}(t)\}$. If $\Vmc_{r}(t) = \emptyset$ then also $\Vmc_{r}^{*}(t) = \emptyset$ is adopted by convention. In addition, we define
\begin{align*}
	\bar{\n}_{r}(t) = \begin{cases}
		\sum_{k \in \Vmc_{r}^{*}(t)}\g_{rk}(t), & \Vmc_{r}^{*}(t) \neq \emptyset; \\
		\g^{*}, & \text{otherwise}.
	\end{cases}
\end{align*}
Its corresponding unit vector version is given by
\begin{equation}\label{eq:nrt}
	\n_{r}(t) = \dfrac{\bar{\n}_{r}(t)}{\left\|\bar{\n}_{r}(t)\right\|}  
\end{equation} 
and is such that $-\n_{r}(t)$ represents the desired escaping direction for the RV. 
Finally, assuming $\gamma \neq \pi/2$, let us also define
\begin{equation}\label{eq:vMbart}
	\bar{v}_{M}(t) =\begin{cases}
		\dfrac{v_{M}}{\n_{r}(t)^{\top}\g_{rk}(t)} &  \text{for any } k \in \Vmc_{r}^{*}(t), \text{ if } \Vmc_{r}^{*}(t) \neq \emptyset; \\
		0, & \text{otherwise};
	\end{cases} 
\end{equation}
denoting the minimum speed required at time $t$ for the RV to avoid collisions while moving virtually along $-\n_{r}(t)$ in the worst case scenario, i.e., when all agents $k \in \Vmc_{r}^{*}(t)$ are moving towards $\p_{r}(t)$ with a velocity $\dot{\p}_{k}(t) = -v_{M} \g_{rk}(t)$. 

It is worth to note that if $\Vmc_{r}^{*}(t) \neq \emptyset$, then $0<\cos_{\gamma} \leq \n_{r}(t)^{\top}\g_{rk}(t) \leq 1$ for all $k \in \Vmc_{r}^{*}(t)$. Hence, $\bar{v}_{M}(t) \in [v_{M},v_{M}/\cos_{\gamma}]$ is bounded for any $\gamma \neq \pi/2$. 
The collision avoidance action required for the RV as soon as $d_{r}(t) = \epsilon$ can be thus identified and quantified as 
\begin{equation}\label{eq:mincaaction}
	(a_{r}(t),\f_{r}(t)) = (\bar{v}_{M}(t), \n_{r}(t)).
\end{equation}
However, such a choice may lead to undesired effects for the FoV maintenance. Hence, to devise a good design approach for $\Bupsilon_{r}(t)$, the following theorem is provided.

\begin{theorem}\label{thm:control_law_general_plus_ca} 
	Given $\gamma \neq \pi/2$, let us consider dynamics \eqref{eq:dynca}-\eqref{eq:etaeps_alertfun} and define $\Vmc_{r}(t)$, $\n_{r}(t)$ and $\bar{v}_{M}(t)$ as in \eqref{eq:Vrset}, \eqref{eq:nrt} and \eqref{eq:vMbart}, respectively.
	If Asm. \ref{asm:coll_avoid_relay} is satisfied at $t=0$, then the set $\Vmc_{\epsilon}(t) = \{i\in \Vmc_{s} ~|~ d_{ri}(t) < \epsilon\}$ remains empty for all $t>0$ by choosing
	\begin{equation}\label{eq:avoid_eff}
		a_{r}(t) = \begin{cases}
			\dfrac{[\bar{v}_{M}(t)+\u_{r}(t)^{\top}\n_{r}(t) ]_{+}}{\n_{r}(t)^{\top}\g^{*}}  &\text{if } \Vmc_{r}(t) \neq \emptyset;\\
			0, & \text{otherwise};
		\end{cases}   
	\end{equation}
	and
	\begin{equation}\label{eq:esc_dir}
		\f_{r}(t) = \g^{*}, \quad \forall t \geq 0,
	\end{equation}
	where $\u_{r}(t)$ is taken as in \eqref{eq:ctrl_law_1} or \eqref{eq:bearing_general_control}. Moreover, Prob. \ref{problem_statement} is solved by using a collision avoidance term $\Bupsilon_{r}(t)$ as in \eqref{eq:collav_factors}, selected according to \eqref{eq:avoid_eff}-\eqref{eq:esc_dir}.
\end{theorem}

\begin{figure}[h!] 
	\centering
	\subfloat[$ w_{r} \geq \bar{v}_{M}$]{\includegraphics[scale=0.25, clip ]{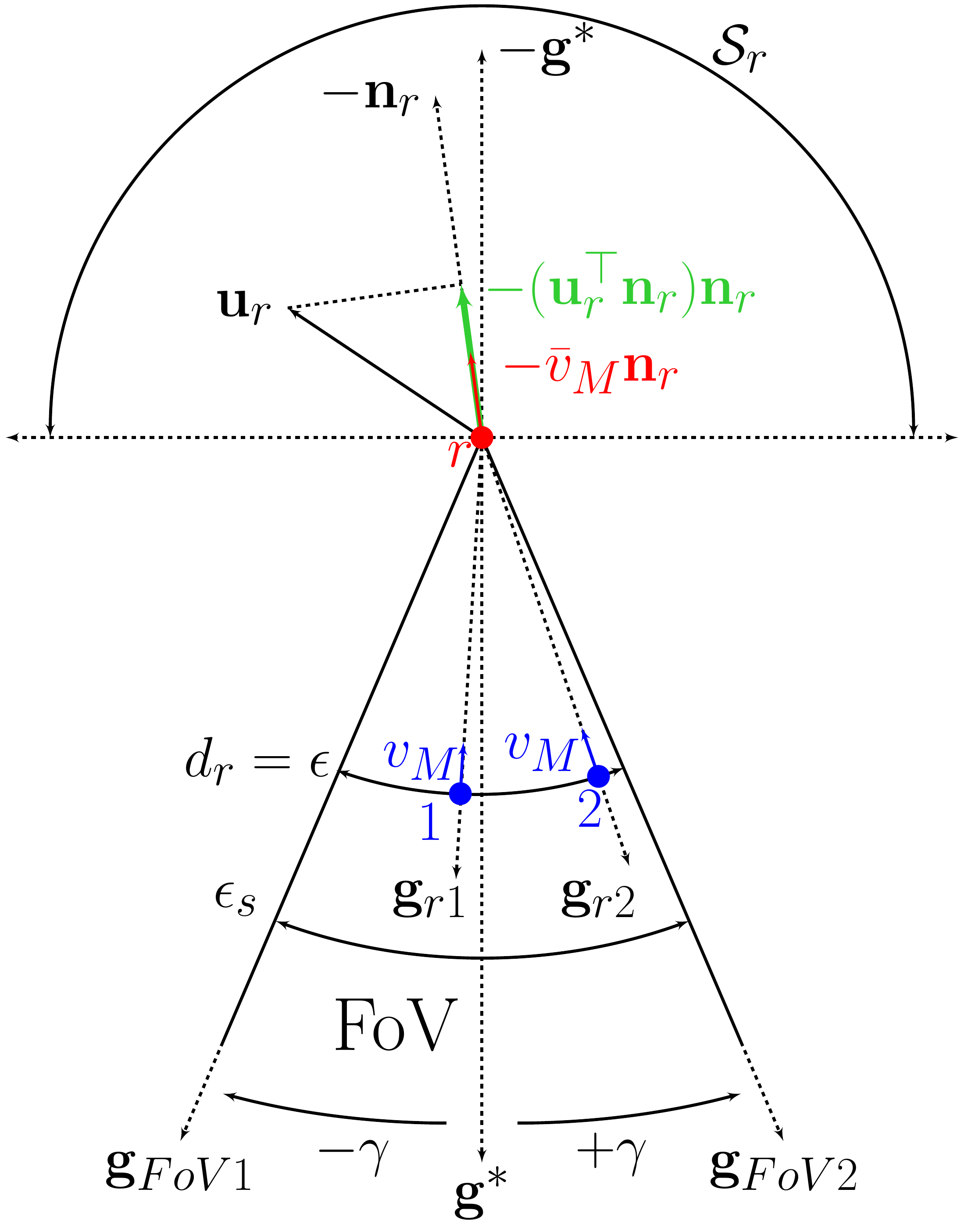}\label{fig:coll_av_repr_w>v}} \hspace{2cm}
	\subfloat[$ w_{r} < \bar{v}_{M}$]{\includegraphics[scale=0.25, clip]{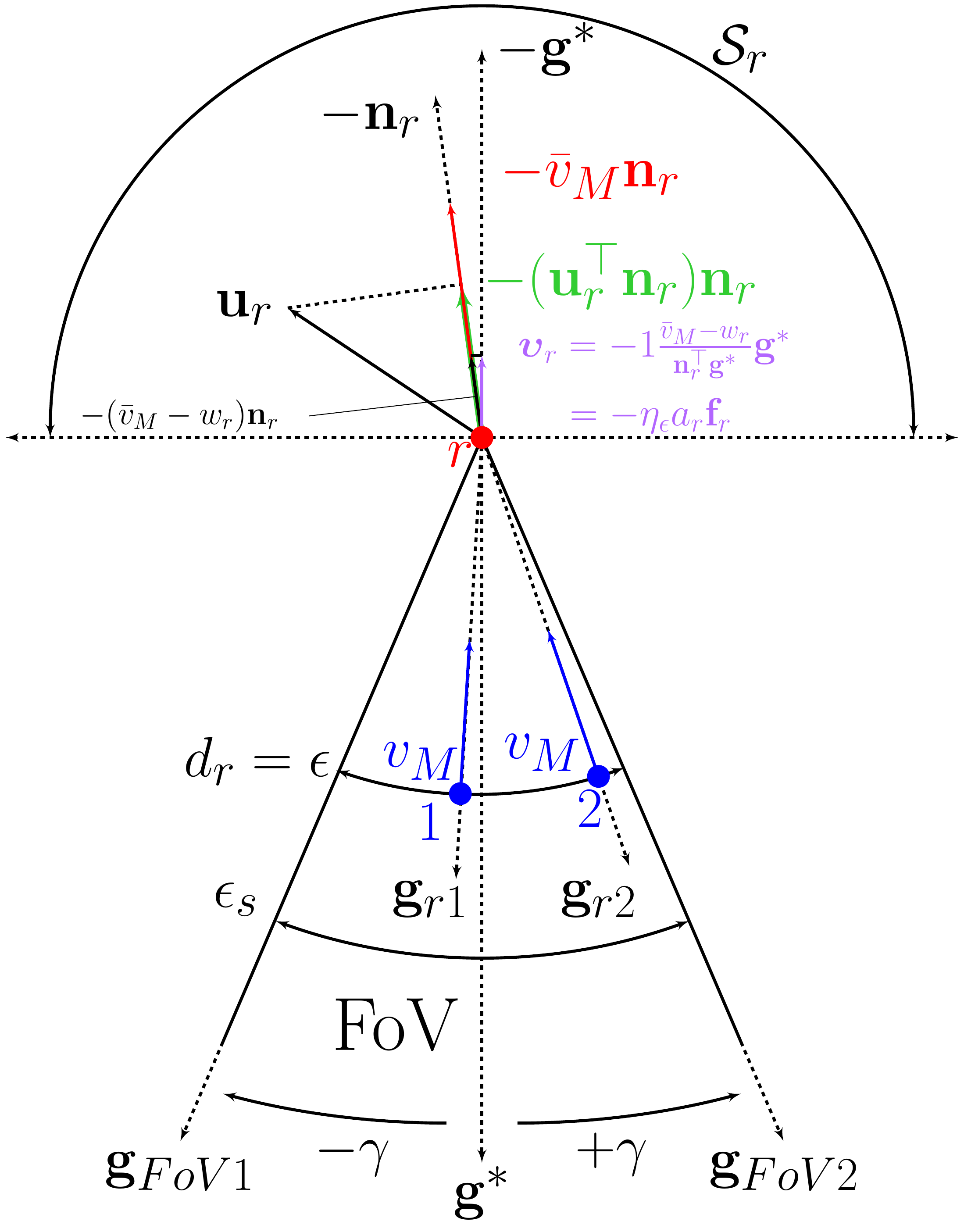}\label{fig:coll_av_repr_w<v}}
	\caption{Illustration of the collision avoidance strategy under analysis: possible scenarios. Here $\Vmc_{r}^{*} = \Vmc_{r} = \{1,2\}$ holds.}
	\label{fig:coll_av_repr}
\end{figure}

\begin{proof}
	Firstly, observe that $\u_{r}(t)^{\top}\g^{*} \leq 0$ holds true for any $\u_{r}(t)$ under consideration and all $t\geq 0$, thus suggesting a characterization for the so-called \textit{admissible motion space} of the RV $\Smc_{r} = \{\Bomega_{r} \in \mathbb{R}^{d} ~|~ \Bomega_{r}\g^{*} \leq 0 \text{ and } \left\|\Bomega_{r} \right\| =1\}$ containing the potential directions $\Bomega_{r}$ for $\dot{\p}_{r}$.
	It is then immediate to verify that $-\g^{*} \in \Smc_{r}$, $\u_{r}(t) \in \Smc_{r}$ and $-\n_{r}(t) \in \Smc_{r}$, $\forall t \geq 0$ by construction. This also denotes that choice in \eqref{eq:esc_dir} is at least admissible to solve Prob. \ref{problem_statement}, thanks to the structure of the collision avoidance term in \eqref{eq:collav_factors} and since $a_{r}(t) $ in \eqref{eq:avoid_eff} is nonnegative for all $t \geq 0$. 
	
	Now, since \eqref{eq:mincaaction} represents the action  for the RV sufficient to avoid collisions in the worst case scenario (without accounting for the control task of $\u_{r}$), assuming that $\eta_{\epsilon}$ activates (i.e. it takes value $1$) for $d_{r} = \epsilon$, in order to show condition $\Vmc_{\epsilon}(t) = \emptyset, \forall t>0$, it is sufficient to ensure 
	\begin{equation}\label{eq:emptyseteqcond}
		\dot{\p}_{r}(t)^{\top} (-\n_{r}(t)) \geq \bar{v}_{M}(t), \quad \forall t>0 .
	\end{equation} 
	Given this premise, the proof can be split into two scenarios resorting to quantity $w_{r}(t) := -\u_{r}^{\top}(t)\n_{r}(t)$ and illustrated by means of Fig. \ref{fig:coll_av_repr}.
	
	$\bullet$ Scenario $w_{r} \geq \bar{v}_{M}$.  
	The latter condition is equivalent to have $\u_{r}^{\top}(-\n_{r}) \geq \bar{v}_{M}$, hence this scenario (depicted in Fig. \ref{fig:coll_av_repr_w>v}) implies that control input $\u_{r}$ itself is sufficient to serve as a collision avoidance action. Indeed, $(a_{r},\f_{r}) = (0,\g^{*})$ guarantees condition \eqref{eq:emptyseteqcond} and solves Prob. \ref{problem_statement}, since $\Bupsilon_{r} = \zeros$ leads to $\dot{\p}_{r}^{\top}(-\n_{r}) = (\u_{r}+\Bupsilon_{r})^{\top}(-\n_{r}) = -\u_{r}^{\top}\n_{r} \geq \bar{v}_{M}$ and it does not affect the control action carried out by $\u_{r}$. As a final remark, notice that the case $\Vmc_{r} = \emptyset$ falls inside this scenario by definition, since $\u_{r}^{\top}\g^{*} \leq 0$ and, in this case, $(\bar{v}_{M},\n_{r})=(0,\g^{*})$.
	
	$\bullet$ Scenario $ w_{r} < \bar{v}_{M}$. In this scenario (depicted in Fig. \ref{fig:coll_av_repr_w<v}), control input $\u_{r}$ is not sufficient to concur fully to a collision avoidance action; hence, additional effort is needed, that is $a_{r}>0$ is required. Such an missing avoidance effort can be identified and quantified as $(v_{M}-w_{r}) > 0$ along the collision direction $-\n_{r}$. Nonetheless, substituting $ -(v_{M}-w_{r}) \n_{r}$ in place of $\Bupsilon_{r} $ and adding it to $\u_{r}$ in \eqref{eq:dynca} may lead to undesired trajectories of $\p_{r}$, as the RV could lose track of the agents inside its FoV. What is however permitted is to exploit the direction of motion given by $-\g^{*}$, as the latter represents an equilibrium for dynamics \eqref{integrator_dynamics}. Indeed, $\u_{r}(\g_{r}) = \zeros$ if all $\g_{r}$ components are equal to $\g^{*}$. In other words, if $\Bupsilon_{r}$ has direction $-\g^{*}$, the FoV control exerted by $\u_{r}$ is not affected. Because of this fact, $\f_{r} = \g^{*}$ is chosen to solve Prob. \ref{problem_statement}. Moreover, $(a_{r},\f_{r}) = ((\bar{v}_{M}-w_{r})/(\n_{r}^{\top}\g^{*}),\g^{*})$ is sufficient to ensure \eqref{eq:emptyseteqcond}; indeed, one has $\dot{\p}_{r}^{\top}(-\n_{r}) = (\u_{r}+\Bupsilon_{r})^{\top}(-\n_{r}) = (\u_{r}-a_{r}\f_{r})^{\top} (-\n_{r}) = -\u_{r}^{\top}\n_{r}+\bar{v}_{M}-w_{r}  = \bar{v}_{M}$. % $\f_{r}$ is set as $\g^{*}$ and the collision avoidance action yielded by $-(\bar{v}_{M}+\u_{r}^{\top}\n_{r}) \n_{r}$ is carried out along direction $-\g^{*}$. 
	%As a consequence, the avoidance effort is here determined by $a_{r} = (\bar{v}_{M}+\u_{r}^{\top}\n_{r}) / (\n_{r}^{\top}\g^{*})$.  
	Also, note that, if $\n_{r}$ is such that $\n_{r}^{\top}\g^{*} \neq 0$ then $a_{r}$ is well defined. This holds true, as $\n_{r}^{\top}\g^{*} \geq \cos_{\gamma} > 0$, assuming that $\gamma \neq \pi / 2$. 
\end{proof}

%%%%%%%%%%%%%%%%%%%%%%%%%%%%%%%%
%%%%%%%%%%%%%%%%%%%%%%%%%%%%%%%%
%%%%%%%%%%%%%%%%%%%%%%%%%%%%%%%%
%\newpage
\section{Numerical simulations}\label{sec:simulations}

To support the theoretical results obtained in Sec. \eqref{sec:systemmodelandcontrolstrategy} we provide here several numerical simulations sharing the following setup. The maximum speed $v_{M} = 5~\si{ms^{-1}}$ is established for all the agents involved (RV excluded) and a FoV angle of $\gamma_{FoV} = 90^{\circ}$ is set%The latter choice is justified by the fact that $\gamma = \pi/4~\si{rad}$ easily allows the exact computation of
, leading to $q_{\pi/4}^{*} = (3\sqrt{2}-2)/4 \simeq 0.5607$. With this setup, we obtain $K_{r}^{\star} \simeq 7.0711~\si{ms^{-1}}$ and $K_{r}^{q} \simeq 8.9181~\si{ms^{-1}}$. According to the previous theoretical results, we define the \textit{critical control gain}\footnote{We also highlight that a tight upper bound of this quantity can be yielded by $K_{rc} \leq \overline{K}_{rc} = v_{M}/\sin_{\gamma}^{3}$ for a generic $\gamma \in (0,\pi/2]$, since, by the structure of \eqref{eq:qgammaofphi}, $q_{\gamma}(\phi)$ is lower bounded by $\sin_{\gamma}^{3}$. Thus, imposing $K_{r} \geq \overline{K}_{rc}$ is sufficient to ensure validity for the proposed control laws for any given admissible value of $\gamma$.} as 
\begin{equation*}
	K_{rc} = \begin{cases}
		K_{r}^{\star} , \quad \text{ if } n = 1; \\
		K_{r}^{q} , \quad \text{ if } n > 1;
	\end{cases}
\end{equation*}
for which the proposed bearing-based control laws \eqref{eq:ctrl_law_1}, \eqref{eq:bearing_general_control} are effective only by adopting a control gain $K_{r}$ such that $K_{r} \geq  K_{rc}$.\\
In addition, we refer to the \textit{FoV borders} as the half lines $\mathbf{h}_{\scaleto{FoVj}{4pt}}(t) = \p_{r}(t)+\lambda \mathbf{g}_{\scaleto{FoVj}{4pt}}$, for $j=1,2$, with $\lambda$ ranging over $[0,+\infty)$. The initial position of the RV is set at $\p_{r}(0) = \zeros_{}$ with a bearing bisector $\g^{*} = \begin{bmatrix}
	0 & -1 
\end{bmatrix}^{\top}$, 
leading to $\mathbf{g}_{\scaleto{FoV1}{4pt}} = \R_{z}(-\gamma) \g^{*}$ and $\mathbf{g}_{\scaleto{FoV2}{4pt}} = \R_{z}(\gamma) \g^{*}$. 

All simulations run over a time interval $T=T_{f}-T_{i}$ ranging from $T_{i} = 0$ to $T_{f}= 30~\si{s}$ and few snap-shots of the trajectories are reported at time instants $t_{k}$, where $k\in \{0,\ldots,5\}$, $t_{0} = T_{i}$ and $t_{k+1} = t_{k}+T/5$. Also, the collision avoidance strategy devised in Sec. \ref{sec:collavoidimpl} is here implemented through $\bar{\eta}_{\epsilon}(d_{r}) = -d_{r}/(\delta \epsilon) + (1+\delta)/\delta$, with $\delta \in (0,1]$, where $\epsilon = 5~\si{m}$ and $\delta = 0.01$ are chosen, given the short-range distance sensing $\epsilon_{s}=2\epsilon$.
In particular, the first group of numerical results asses the validity and limitations of the proposed control gain selection; whereas, the second one testes the switching mechanism for $n>1$ agents. Lastly, the final paragraph is devoted to a potential real-world application example.

\subsection{Validation of the control gain selection and limitations}
In this framework, we show how the control gain selection influences the RV's trajectory and the maintenance of either a single agent or a couple of agents under its tracking action.

\subsubsection{Single agent case}\label{ssec:singleagentcaseval}
\begin{figure*}[t!] 
	\centering
	\hspace{-0.65cm}
	\subfloat[$K_{r} = 0.9 K_{rc}$]{\includegraphics[width=0.25\textwidth, trim={11cm 0cm 11cm 0cm},clip ]{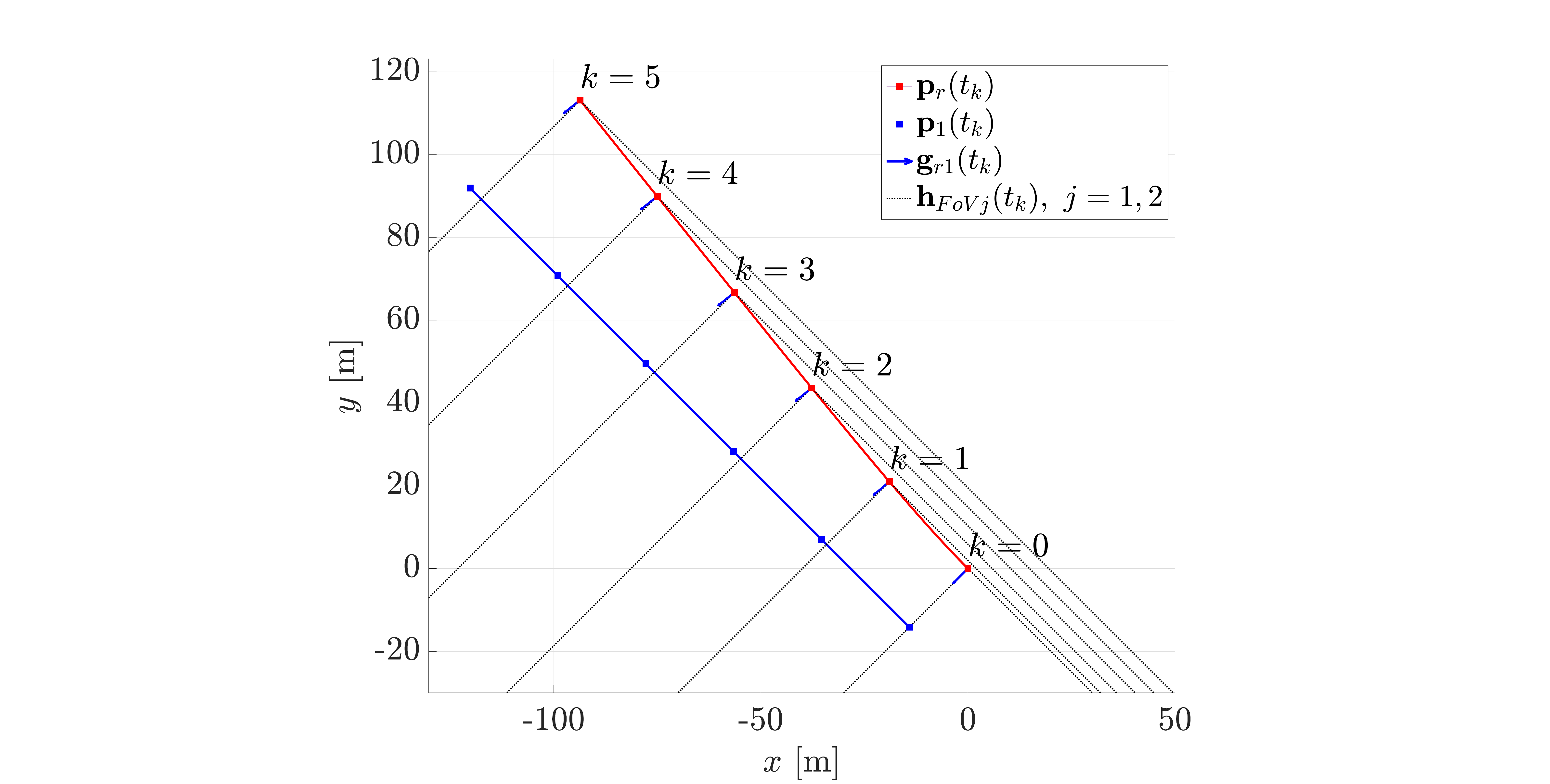}\label{fig:sac0_9}}
	\subfloat[$K_{r} = K_{rc}$]{\includegraphics[width=0.25\textwidth, trim={11cm 0cm 11cm 0cm},clip]{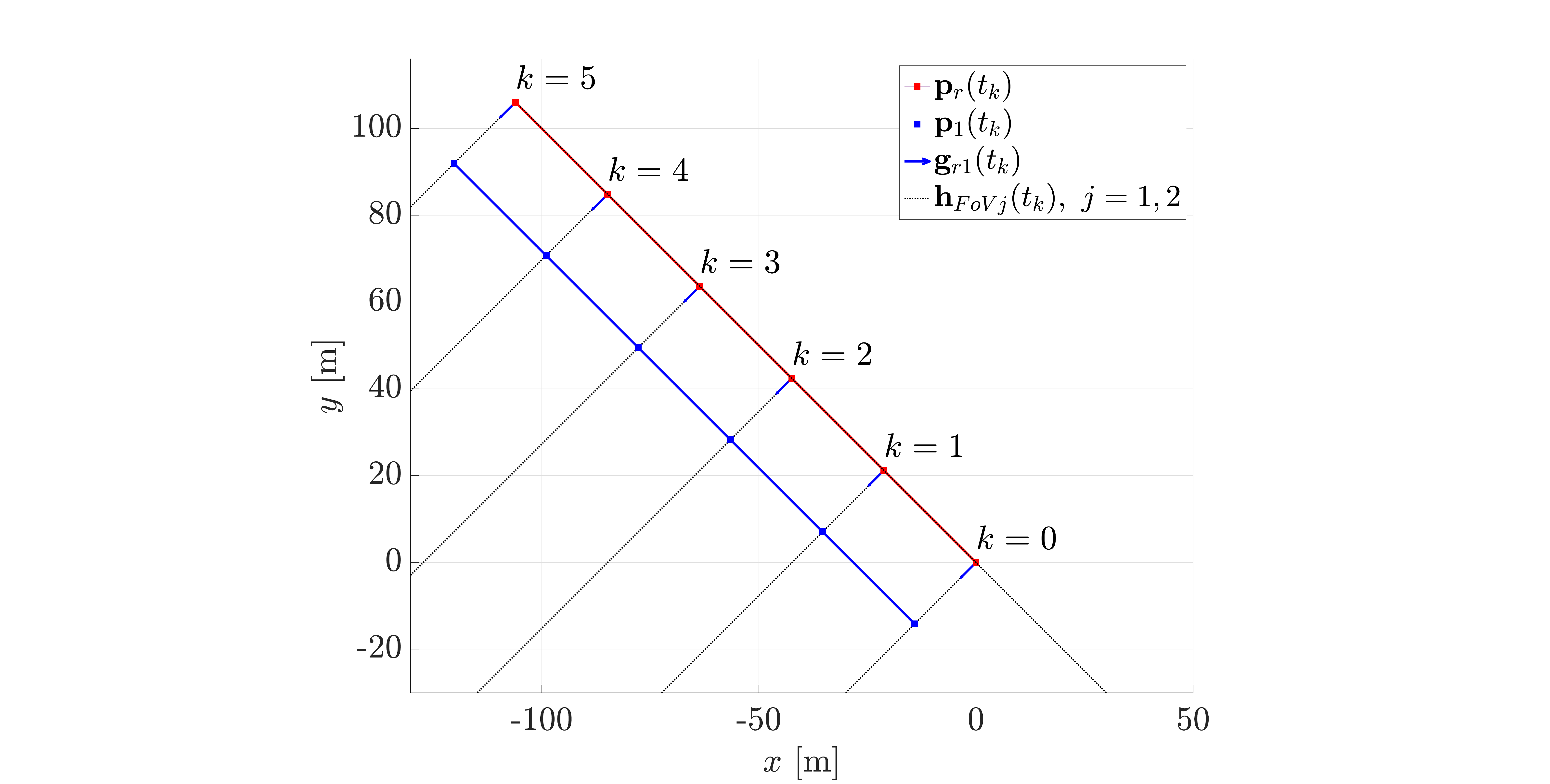}\label{fig:sac1}}
	\subfloat[$K_{r} = 1.1 K_{rc}$]{\includegraphics[width=0.25\textwidth, trim={11cm 0cm 11cm 0cm},clip]{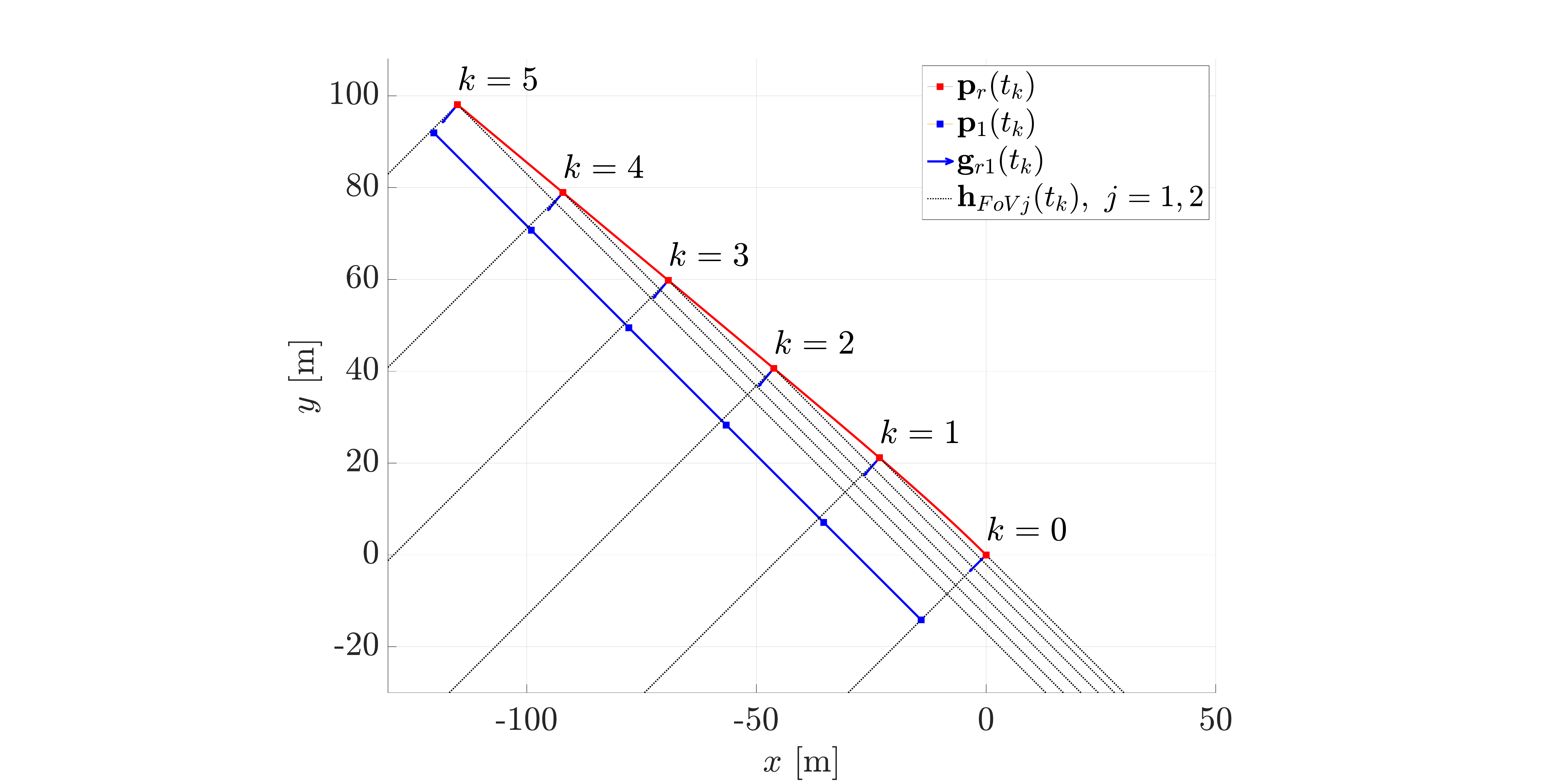}\label{fig:sac1_1}}
	\subfloat[$K_{r} = 1.5 K_{rc}$]{\includegraphics[width=0.25\textwidth, trim={11cm 0cm 11cm 0cm},clip]{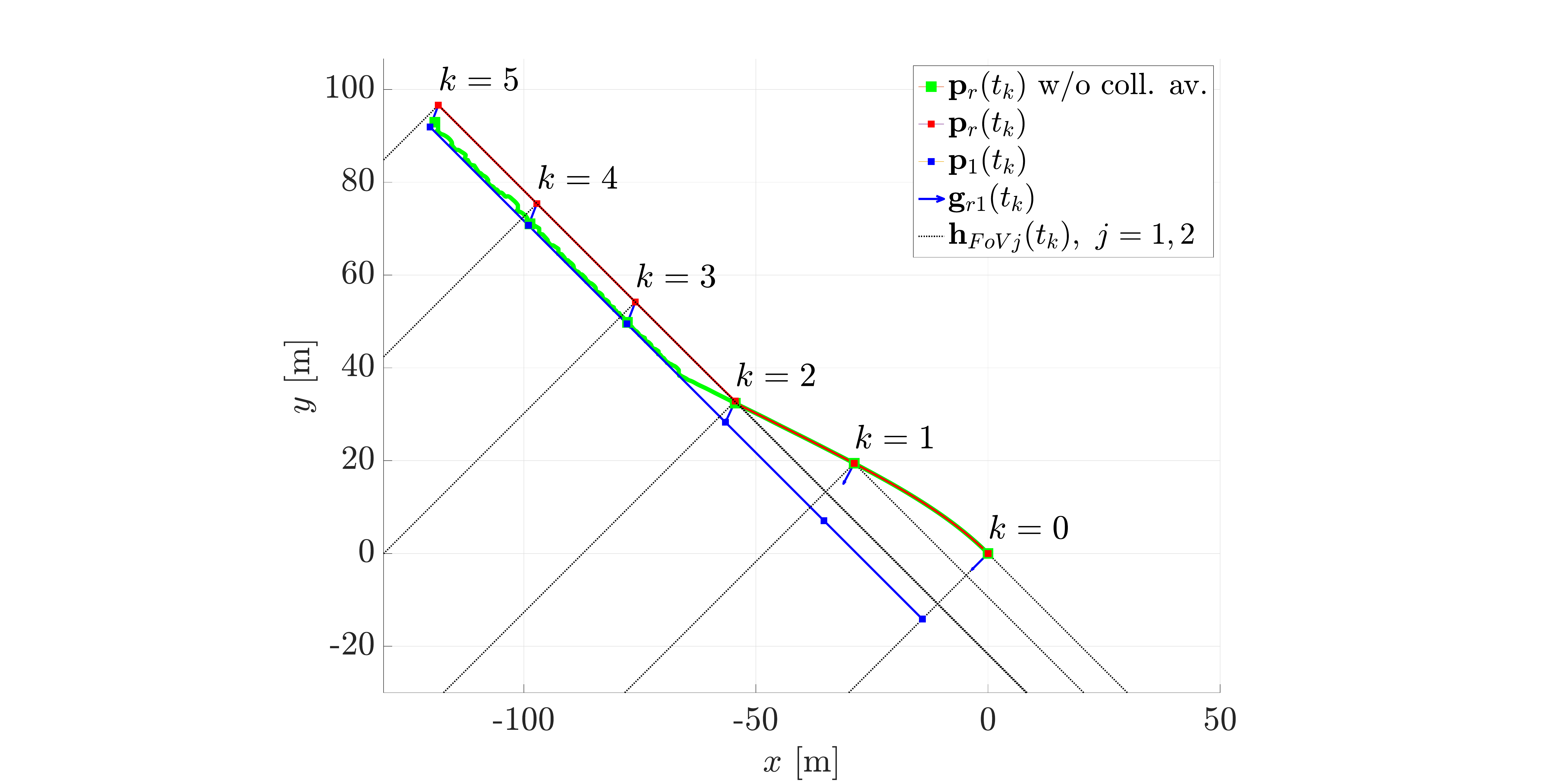}\label{fig:sac1_5}}
	\caption{Gain selection in the single agent case: few examples referring to the worst case scenario described in Prop. \ref{prop:oneagent}}
	\label{fig:gainselection1}
\end{figure*}
\begin{figure}[t!] 
	\centering
	\subfloat[$K_{r} = 0.9 K_{rc}$]{\includegraphics[scale=0.14, trim={5cm 0.2cm 5cm 1.8cm}, clip ]{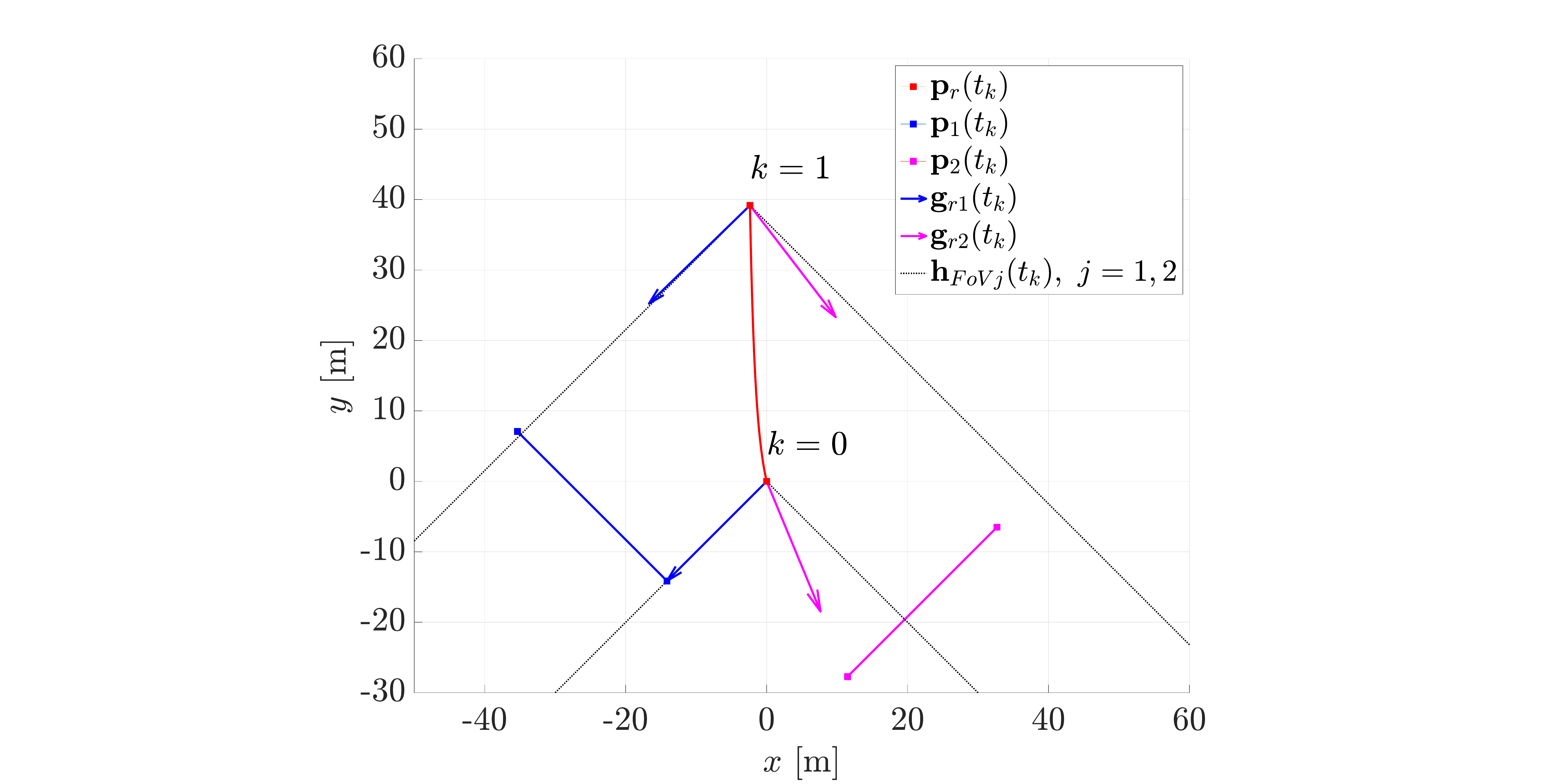}\label{fig:tac0_9}} 
	\subfloat[$K_{r} = K_{rc}$]{\includegraphics[scale=0.14, trim={5cm 0.2cm 5cm 1.8cm}, clip]{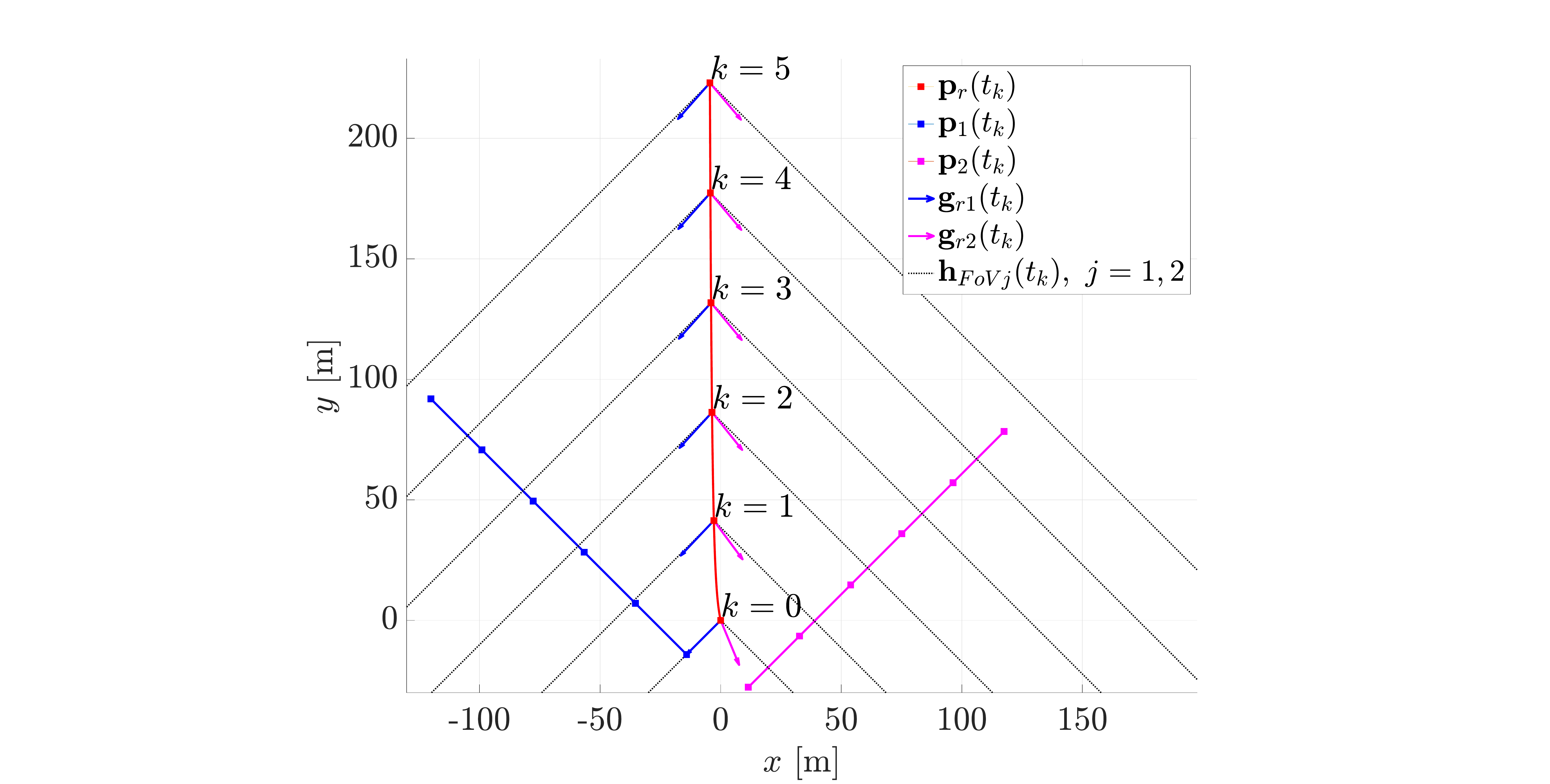}\label{fig:tac1}}
	\caption{Gain selection in the two agent case: few examples referring to the worst case scenario described in Lem. \ref{lemma:twoagents}}
	\label{fig:gainselection2}
\end{figure}
With this group of simulations we intend to support the control gain selection proposed in Prop. \ref{prop:oneagent} when the single agent being tracked is escaping the FoV border $\mathbf{h}_{\scaleto{FoV1}{4pt}}$ and traveling in a straight path with a constant velocity $\dot{\p}_{1}$ perpendicular to $\mathbf{h}_{\scaleto{FoV1}{4pt}}$, such that $\left\|\dot{\p}_{1}\right\| = v_{M}$. Fig. \ref{fig:gainselection1} illustrates the behavior of the RV for different choice of the control gain $K_{r}$ w.r.t. the critical gain $K_{rc}$. In particular, it is possible to appreciate that a gain $K_{r} = 0.9K_{rc}$ (see Fig. \ref{fig:sac0_9}) is insufficient to maintain the tracking, as the agent exits the FoV as soon as the simulation starts. From Fig. \ref{fig:sac1}, it is instead clear that the RV preserves the agent tracking precisely on FoV frontier since $t=0$, if $K_{r}=K_{rc}$ is set. Fig. \ref{fig:sac1_1} depicts the RV's trajectory for $K_{r} = 1.1 K_{rc}$: here, as time instants grow, it is possible to observe that the bearing $\g_{r1}$ points more and more inward w.r.t. the FoV, that is, for $t\rightarrow\infty$, $\g_{r1}$ aligns with $\g^{*}$ because of the structure of law \eqref{eq:ctrl_law_1}. Lastly, Fig. \ref{fig:sac1_5}, wherein $K_{r} = 1.5 K_{rc}$ is taken, shows how essential a collision avoidance strategy is in order to allow the RV not to crash against agents. Indeed, thanks to the method provided in Thm. \ref{thm:control_law_general_plus_ca}, distance $d_{r1}(t)$ remains greater than $5.0288~\si{m} \geq \epsilon$ for all $t\geq0$; whereas, $d_{r1}(t)$ would approach $0$ as $t$ grows, if $\Bupsilon_{r}(t) = \zeros_{}$ were assigned (see trajectory in green).

\subsubsection{Two agent case}
Within this subframework, we aim to justify the gain selection discussed in Lem. \ref{lemma:twoagents}. The setup here adopted adheres to the nontrivial worst case scenario arising from said lemma: at $t=0$, we set agent $1$ on the FoV border $\mathbf{h}_{\scaleto{FoV1}{4pt}}$ and agent $2$ close to the other border, describing an angle $-\phi^{*}_{\gamma}$ from $\mathbf{h}_{\scaleto{FoV2}{4pt}}$, where $\phi^{*}_{\pi/4}=\pi/8~\si{rad}$ is the angle minimizing $q_{\gamma}(\phi)$, i.e. $q_{\gamma}(\phi^{*}_{\gamma}) = q_{\gamma}^{*}$. The numerical simulations are reported in Fig. \ref{fig:gainselection2}. In particular, the selection $K_{r}=0.9 K_{rc}$ in Fig. \ref{fig:tac0_9} leads to a relay tracking failure, starting from the very first time instants (see instant $t_{1}$, in which agent $1$ clearly exits the FoV). On the contrary, Fig. \ref{fig:tac1} describes the presence of sufficient capabilities for the RV to maintain both agents inside the FoV over the entire time interval $T$.

%\pagebreak

\begin{figure}[t!]
	\centering
	%\hspace{-1.7cm}
	\includegraphics[scale=0.17,trim={11cm 0.2cm 11cm 0.4cm},clip]{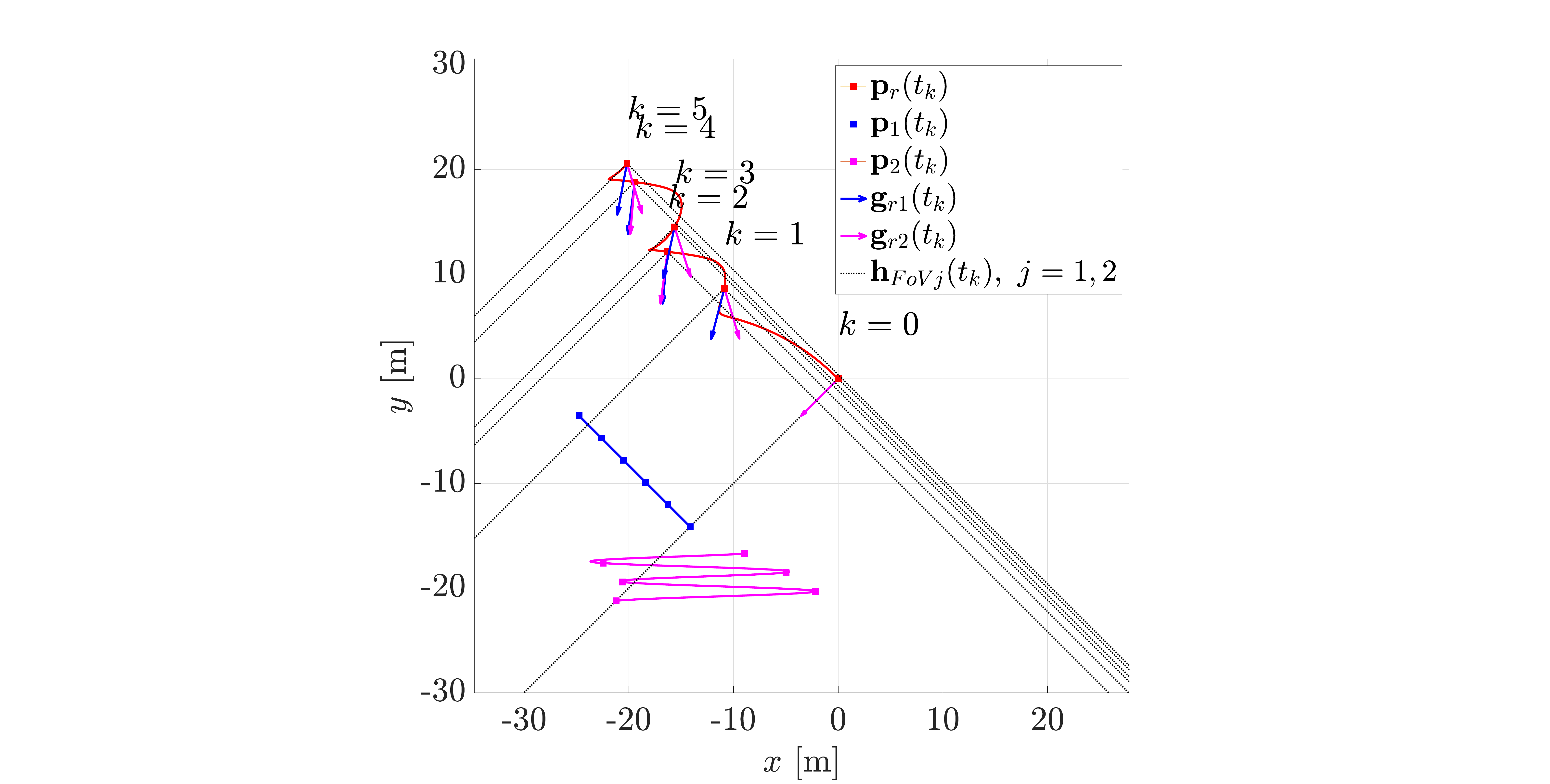}
	\caption{The continuity of law \eqref{eq:ctrl_law_2} is tested while its action switches over time. Two agents are monitored by the RV.}
	\label{fig:2dancingagents}
\end{figure}

\subsection{Validation of the switching mechanism}
One of the main concerns about control laws \eqref{eq:ctrl_law_2} and \eqref{eq:bearing_general_control} is represented by the manifestation of chattering phenomena, in practice, when the controller switches from one policy to the other (e.g. from case $\chi_{n} \geq 0$ to case $\chi_{n} < 0$ and vice-versa). In the following lines we provide few evidences showing that this issue does not subsist.
In both simulations $K_{r} = K_{rc}$ is assumed and at least one of the agents is characterized by a non linear dynamics designed ad-hoc to track the bisector direction $\g^{*}$ with possibility of overshoot.

\subsubsection{Two agent case}
In this example, we further reinforce the validity of Lem. \ref{lemma:twoagents} by showing that the switching behavior of the controller does not affect negatively the tracking performances. Fig. \ref{fig:2dancingagents} illustrates this simulation: agent $2$ crosses the bisector direction $5$ times over interval $T$ and causes the sign changes of $\chi_{2}$, leading to the same number of switches for the control policy in \eqref{eq:ctrl_law_2}.

\subsubsection{Generalization for $n\geq 1$ agents}
\begin{figure}[!t]
	\centering
	%\hspace{-0.9cm}
	\includegraphics[scale=0.16,trim={11cm 0.2cm 11cm 0.4cm},clip]{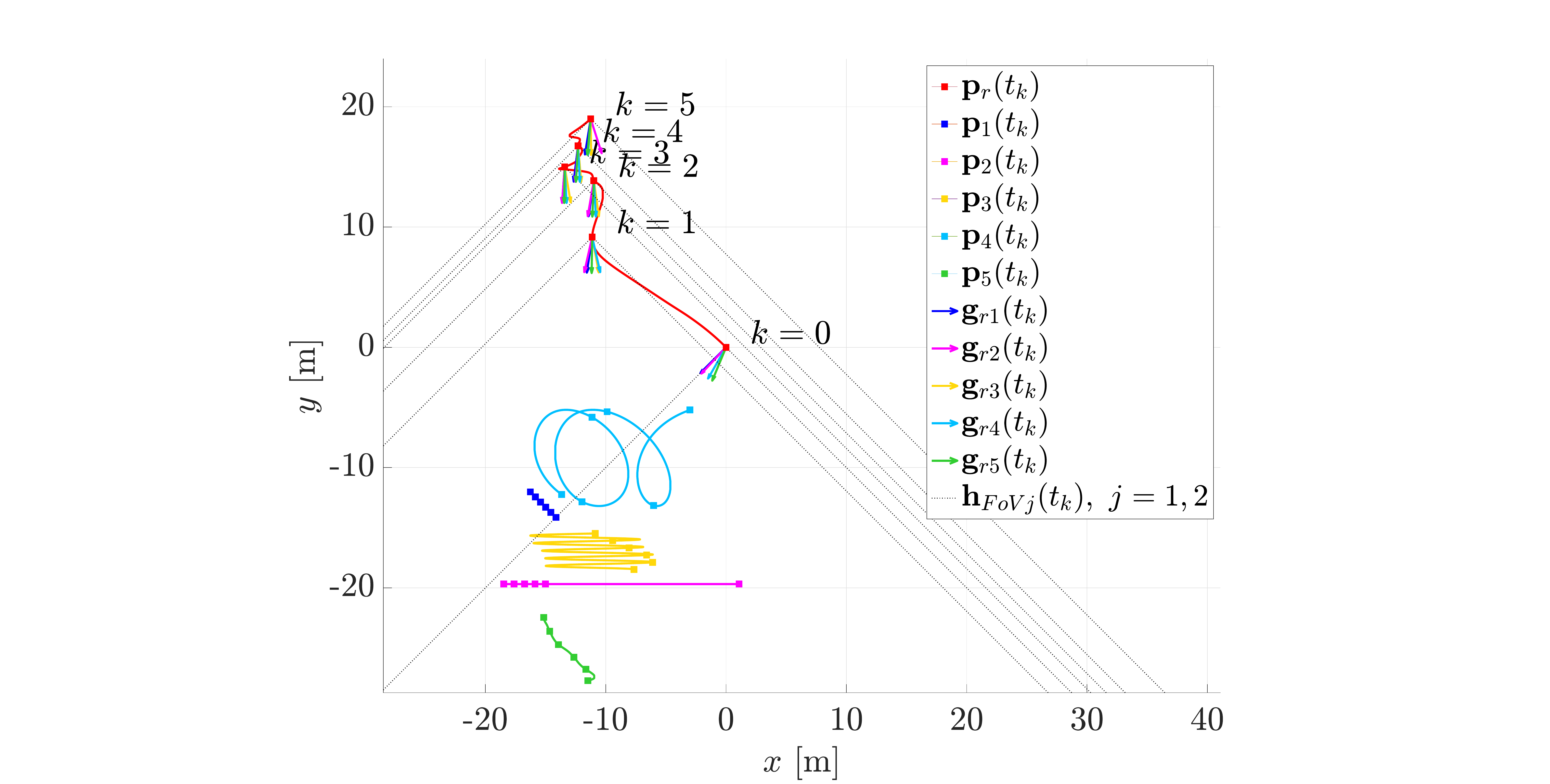}
	\caption{The continuity of law \eqref{eq:bearing_general_control} is tested while its action switches over time. Five agents are monitored by the RV.}
	\label{fig:5dancingagents}
\end{figure}

To support Thm. \ref{thm:control_law_general} and the switching capability of law \eqref{eq:bearing_general_control} we have designed a numerical simulation involving $n=5$ agents. In Fig. \ref{fig:5dancingagents} the relative tracking performances are depicted. It is worth to observe that not only $\chi_{n}$ changes sign $3$ times, leading to the same number of switches for the control policy in \eqref{eq:bearing_general_control}, but also that maximization in \eqref{eq:grjbar} yields several different results over the interval $T$. In other words, it is possible to appreciate that the bearing vectors on which the control action within a specific policy of \eqref{eq:bearing_general_control} is computed depends on the closest agents to the FoV borders $\mathbf{h}_{\scaleto{FoVj}{4pt}}(t)$, $j=1,2$, at each time instant $t$.

\begin{figure}[!t]
	\centering
	%\hspace{-1.9cm}
	% trim={<left> <lower> <right> <upper>}
	\includegraphics[scale=0.17, trim={11cm 0.2cm 11cm 0.4cm}, clip]{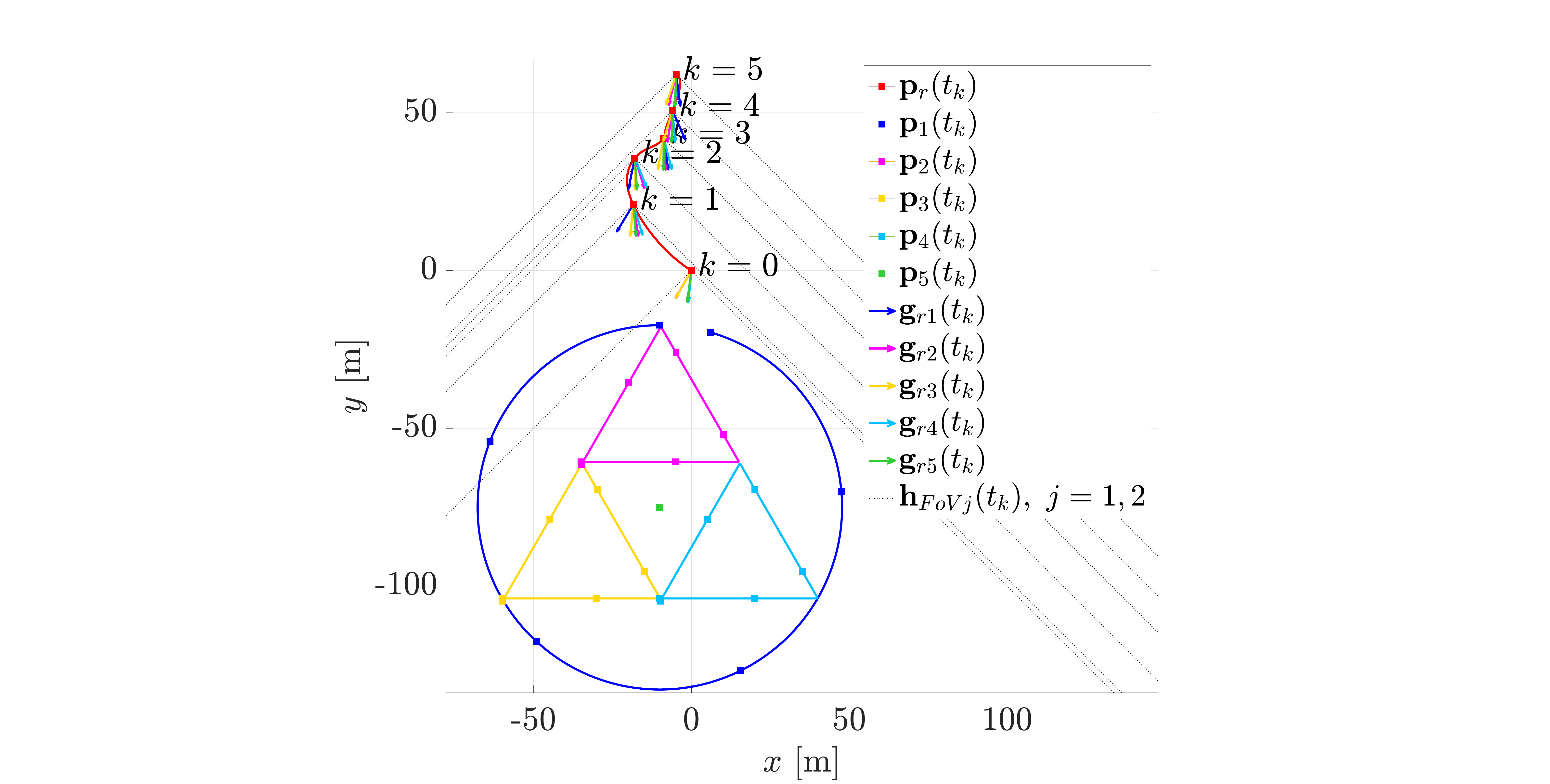}
	\caption{Application example based on a surveillance-type scenario involving approaches from patrolling and dynamic coverage.
	}
	\label{fig:5nagents_patrolling}
\end{figure}

\subsection{Application example}
With this final paragraph we intend to provide an application example giving more credit to the key theoretical result in this paper, namely Thm. \ref{thm:control_law_general}. Specifically, the following numerical simulation pertains to typical aspects revolving around \textit{patrolling} (see e.g. works on camera network patrolling under unreliable communication \cite{BofCarliCenedese2017}, distributed partitioning strategies, visual optimization and perimeter patrolling \cite{BelgioiosoCenedeseMichieletto2016}) and \textit{dynamic coverage} (see e.g. works on sensor networks \cite{LiuDousseNain2013}, UAVs in a cellular system \cite{LiCai2017}). The common approach underneath these research fields usually leverages optimization results on objects moving along a linear path.

Frequently, such studies lead to cope with regular spacial shapes and control patterns in order to govern optimally the trajectories of the objects in question. At the light of this preliminary consideration, we propose here a surveillance-type scenario in which the RV monitors a camera network made up by $n=5$ elements, each of them employed to cover/sense a certain portion of the planar environment. As illustrated in Fig. \ref{fig:5nagents_patrolling}, we let camera $1$ steer along a circular path (in blue), cameras $2,3,4$ move along triangular patterns (in magenta, yellow and cyan) within the aforementioned circular path and camera $5$ stand still at the center (in green) of the circle.

Again, we have set $K_{r} = K_{rc}$ to allow full tracking capabilities for the RV, as it is shown. From the simulation data, it is also possible to observe that controller \eqref{eq:bearing_general_control} switches its regime once, from $\chi_{n}=1$ to $\chi_{n}=-1$ at $t=t_{s}=1.0870~\si{s}$. Indeed, for $t < t_{s}$, one has $\chi_{n}=1$; thus, all agents are identified as geometric elements laying on the ``same side'' of the FoV (the left one, precisely) and camera $1$ is considered the most external agent that can potentially exit the FoV border $\mathbf{h}_{\scaleto{FoV1}{4pt}}(t)$. Moreover, cameras $1,3,4$ are the most recurrent agents determining control action $\chi_{n}=-1$ computed in \eqref{eq:bearing_general_control}, as $t \geq t_{s}$ (all cameras are recognized to belong to ``different sides'' of the FoV within this regime).

%%%%%%%%%%%%%%%%%%%%%%%%%%%%%%
%%%%%%%%%%%%%%%%%%%%%%%%%%%%%%
%%%%%%%%%%%%%%%%%%%%%%%%%%%%%%

\section{Conclusions and continued research}\label{sec:conclusions}
In this paper, we have tackled the problem of communication relay establishment for multiple mobile vehicles by leveraging well-known formation control techniques. The bearing-based strategy devised allows the design of a scalable distributed control law that accounts for FoV constraints that, remarkably, introduce hard nonlinearities to the system of agents under consideration. The proposed bearing-based control law is also endowed with a collision avoidance strategy that employs short-range distance measurements, granting the RV to maintain the agents under its FoV while preventing physical impacts from occurring. The numerical simulations and application examples reported strongly match the theoretical results and performance analysis of the underlying approach. An extension to the three dimensional environments ($d=3$) is envisaged as future work.

\section*{Acknowledgements}

We would like to express appreciation to Nathaniel Drellich for his valuable and constructive suggestions during the planning and development of this research work.

\appendix 
\section{Analysis of function \lowercase{$q_{\gamma}(\phi)$}}\label{sec:appendix}

\begin{figure}[!b]
	\centering
	\includegraphics[scale=0.28]{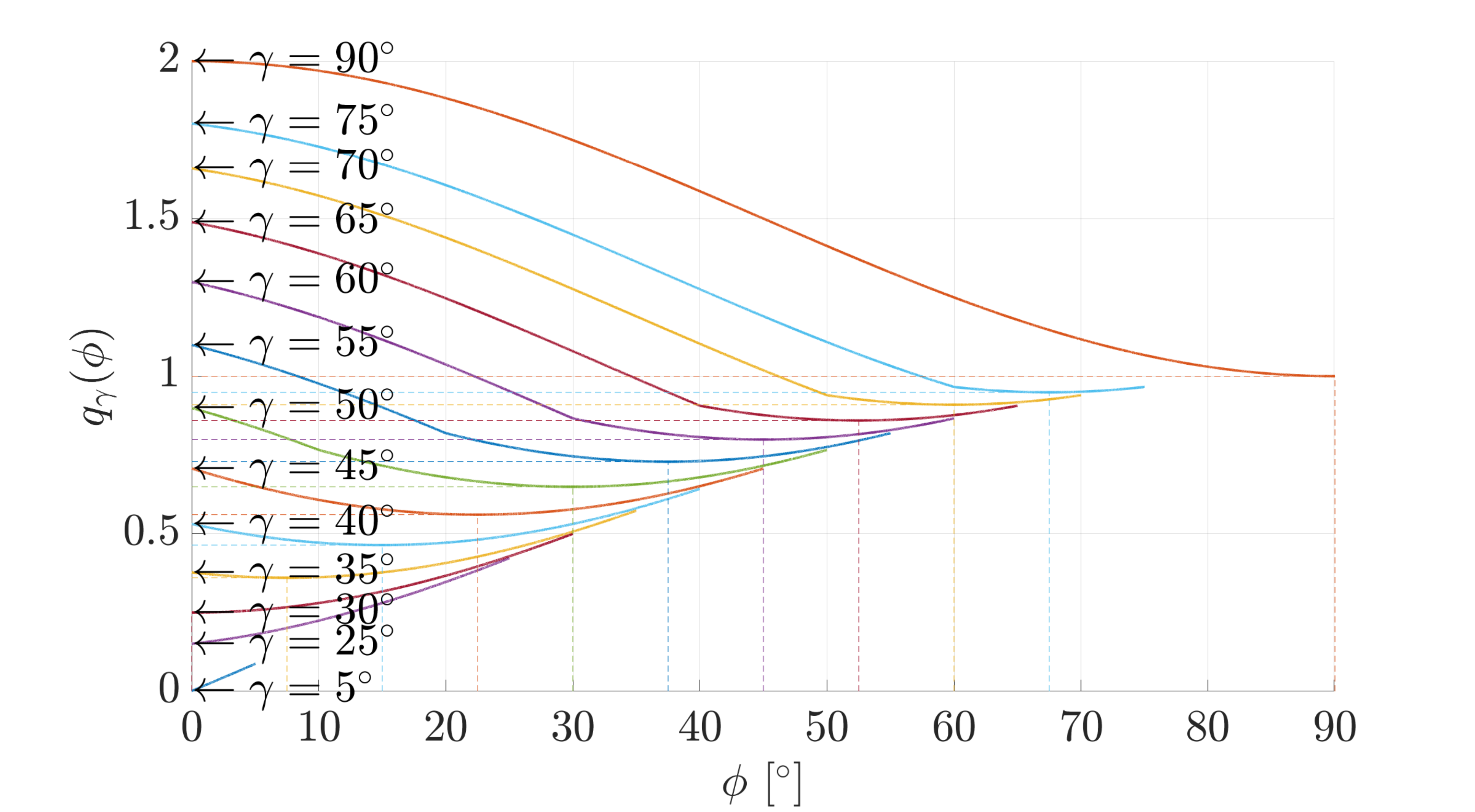}
	\caption{Plot of function $q_{\gamma}(\phi)$ for some values of $\gamma \in (0, \pi/2 ] $ as $\phi$ varies in $[0,\gamma]$. Dashed lines indicate minima $(\phi^{*}_{\gamma},q_{\gamma}^{*})$.}
	\label{fig:qgammaphi}
\end{figure}

Here, the extrema of the function $q_{\gamma}(\phi)$ introduced in \eqref{eq:qgammaofphi} are investigated formally. A plot of $q_{\gamma}(\phi)$ for different values of $\gamma$ is also available in Fig. \ref{fig:qgammaphi}.

$~$\\
\begin{proposition}\label{prop:minimizerq}
		Let $\gamma$ belong to $(0, \pi/2] $. Then there exists a unique global minimizer $\phi^{*}_{\gamma}$ for the function $q_{\gamma}(\phi)$ defined in \eqref{eq:qgammaofphi}, given as
		\begin{equation}\label{eq:minimizer_expression}
			\phi_{\gamma}^{*} = \underset{\phi \in [0,\gamma]}{\arg \min} ~ q_\gamma(\phi) = \begin{cases}
				0, &\quad \text{if } \gamma \in \left(  0,\frac{\pi}{6} \right]; \\
				\frac{3}{2}\gamma-\frac{\pi}{4}, &\quad \text{if } \gamma \in \left( \frac{\pi}{6}, \frac{\pi}{2} \right].
			\end{cases}
		\end{equation}
		Moreover, if $\gamma \in (0 , \pi/6  ) $ then $q_{\gamma}(\phi)$ has no stationary points. Whereas, if $\gamma \in [\pi/6 , \pi/2 ) $ then $q_{\gamma}(\phi)$ has a unique stationary point coinciding with global minimum $\phi^{*}_{\gamma}$. Lastly, if $\gamma = \pi/2$ then $q_{\gamma}(\phi)$ has two stationary points: one is a global maximum, i.e. $\phi = 0$, and the other is a global minimum, i.e. $\phi = \phi^{*}_{\pi/2}$.
\end{proposition}

\begin{proof}
	The proof is split into two parts: the trivial case $\gamma = \pi/2$ and the more articulated case $\gamma \in (0,\pi/2) $.
	
	For $\gamma = \pi/2$, the function \eqref{eq:qgammaofphi} can be easily simplified to
		\begin{equation}\label{eq:q_g_simpl_halfpi}
			q_{\pi/2}(\phi) = 1+ \cos_{\phi}^{2}.
		\end{equation}
		Computing the first derivative $q_{\pi/2}^{\prime}(\phi)$ of \eqref{eq:q_g_simpl_halfpi} w.r.t. $\phi$ one obtains 
		\begin{equation*}
			q_{\pi/2}^{\prime}(\phi) = -\sin_{2\phi},
		\end{equation*}
		which is equal to zero for $\phi = 0$ and $\phi = \phi^{*}_{\pi/2} = \pi/2$. Then, computing the second derivative $q_{\pi/2}^{\prime\prime}(\phi)$ of \eqref{eq:q_g_simpl_halfpi} w.r.t. $\phi$ one has
		\begin{equation*}
			q_{\pi/2}^{\prime\prime}(\phi) = -2\cos_{2\phi},
		\end{equation*}
		which is negative for $\phi = 0$, as $q_{\pi/2}^{\prime\prime}(0) = -2$, and positive for $\phi = \pi/2$, as $q_{\pi/2}^{\prime\prime}(\pi/{2}) = 2$. Hence, there exists a unique global minimizer $\phi^{*}_{\pi/2} = \pi/2$ for $q_{\pi/2}(\phi)$. Also, $q_{\pi/2}(\phi)$ has an additional stationary point that is a global maximum (i.e. at $\phi = 0$).

	We now consider $\gamma \in (0,\pi/2 ) $. By resorting to identity $\sin_{\bar{\phi}}^{2} \cos_{2\bar{\gamma}-\bar{\phi}}^{2} = \sin_{\bar{\gamma}}^{2}+\sin_{\bar{\gamma}-\bar{\phi}}^{2}-2\cos_{2\bar{\gamma}-\bar{\phi}}\sin_{\bar{\gamma}}\sin_{\bar{\gamma}-\bar{\phi}}-(\sin_{\bar{\gamma}}^{2}+\sin_{\bar{\gamma}-\bar{\phi}}^{2})^{2}$,  valid for all $\bar{\gamma}, \bar{\phi} \in \mathbb{R}$, the squared root term in \eqref{eq:qgammaofphi} can be simplified yielding 
		\begin{equation}\label{eq:q_g_simpl}
			q_{\gamma}(\phi) = \sin_{\gamma}^{3} + \sin_{\gamma} \sin_{\gamma-\phi}^{2} + \cos_{\gamma} \sin_{\phi} |\cos_{2\gamma-\phi}| .
		\end{equation}
		Consequently, the first derivative $q_{\gamma}^{\prime}(\phi)$ w.r.t. $\phi$ of \eqref{eq:q_g_simpl} can be now computed and analyzed more easily. It turns out that its expression is well-defined for all $\phi \neq 2\gamma - \pi/2$, whenever $\gamma \in [\pi/4,\pi/2 ) $, and is given by this compact form:
		\begin{equation}\label{eq:q_g_simpl_prime}
			q_{\gamma}^{\prime}(\phi) = \mathrm{sign}(\cos_{2\gamma-\phi}) \cos_{\gamma} \cos_{2(\gamma-\phi)}-\sin_{\gamma} \sin_{2(\gamma-\phi)} .
		\end{equation}
		If $(\gamma,\phi) \in [\pi/4,\pi/2 ) \times [0,2\gamma-\pi/2 )$ then \eqref{eq:q_g_simpl_prime} becomes
		\begin{equation}\label{eq:q_g_simpl_prime_neg}
			q_{\gamma}^{\prime}(\phi)  = -\cos_{\gamma-2\phi} < 0,
		\end{equation}
		since $(\gamma-2\phi) \in (-3\gamma+\pi/2,\gamma ) \subseteq (-\gamma, \gamma ) \subseteq (-\pi/2,\pi/2 )$.
		Whereas, if $(\gamma,\phi) \in [\pi/4,\pi/2 ) \times (2\gamma-\pi/2, \gamma )$ or $(\gamma,\phi) \in ( 0,\pi/4 ) \times [0,\gamma]$ then \eqref{eq:q_g_simpl_prime} becomes
		\begin{equation}\label{eq:q_g_simpl_prime_pos}
			q_{\gamma}^{\prime}(\phi)= \cos_{3\gamma-2\phi}.
		\end{equation}
		First derivative \eqref{eq:q_g_simpl_prime_pos} is null if and only if
		\begin{equation}\label{eq:minimum_candidates}
			\phi = \frac{3}{2}\gamma-\frac{\pi}{4}, \quad \text{for }  \gamma \in \left[\frac{\pi}{6}, \frac{\pi}{2} \right)   ;
		\end{equation}
		therefore, values of $\phi$ in \eqref{eq:minimum_candidates} represent all candidate stationary points for \eqref{eq:q_g_simpl} as $\gamma$ varies. Note that, for $\gamma \in (0,\pi/6)$, no stationary point exists because $\phi \geq 0$ needs to hold by definition.
		
	Next, we provide the expression of the second derivative $q^{\prime \prime}_{\gamma}(\phi)$ w.r.t. $\phi$ of \eqref{eq:q_g_simpl} in order to show the strict convexity of \eqref{eq:q_g_simpl} over the intervals $(\gamma, \phi) \in [\pi/4,\pi/2 ) \times (2\gamma-\pi/2, \gamma )$ and $(\gamma,\phi) \in ( 0,\pi/4 ) \times [0,\gamma]$. In particular, over such intervals, one has
		\begin{equation}\label{eq:q_g_simpl_second_pos}
			q_{\gamma}^{\prime\prime}(\phi) = 2 \sin_{3\gamma-2\phi} > 0,
		\end{equation}
		which is a positive quantity since $0 < 3\gamma - 2\phi < \pi$ holds in this nested case. Inequality \eqref{eq:q_g_simpl_second_pos} thus implies that all values of $\phi$ in \eqref{eq:minimum_candidates} are (at least local) minima.
	
	To conclude the proof, it is worth to observe that $q_{\gamma}(\phi)$ is continuous in $\phi$ for all values of $\gamma \in (0, \pi/2 ) $. This fact and inequalities \eqref{eq:q_g_simpl_prime_neg}, \eqref{eq:q_g_simpl_second_pos} lead to the following implications. On one hand, if $\gamma \in ( 0,\pi/6 ] $, function $q_{\gamma}(\phi)$ is strictly increasing for all $\phi \in [0,\gamma]$. On the other hand, if $\gamma \in ( \pi/6, \pi/2 ) $, function $q_{\gamma}(\phi)$ is strictly decreasing over the interval $[0, 3/(2\gamma)-\pi/4 ] $ and strictly increasing over the interval $[3/(2\gamma)-\pi/4, \gamma ] $. Hence, there exists a global unique minimizer $\phi^{*}_{\gamma}$ for $q_{\gamma}(\phi)$, whose overall expression is given in \eqref{eq:minimizer_expression}. 
\end{proof}

\begin{corollary}
		Under the same assumptions of Prop. \ref{prop:minimizerq}, the minimum value $q_{\gamma}^{*}$ attained by $q_{\gamma}(\phi)$ is given by \eqref{eq:qstarvalue} and $\max(2\sin_{\gamma}^{3},\sin_{\gamma})$ is the maximum value attained by $q_{\gamma}(\phi)$.
\end{corollary}
\begin{proof}
	The minimization follows from the computation of $q_{\gamma}^{*} = q_{\gamma}(\phi^{*}_{\gamma})$ through \eqref{eq:q_g_simpl_halfpi} and \eqref{eq:q_g_simpl}, wherein expression of $\phi^{*}_{\gamma}$ is yielded by \eqref{eq:minimizer_expression}. Similarly, leveraging the Weierstrass theorem and the observations made in Prop. \ref{prop:minimizerq}, $q_{\gamma}(\phi)$ is maximized at $\phi = \gamma$, if $\gamma \in ( 0, \pi/4] $ and at $\phi = 0$, if $\gamma \in [ \pi/4, \pi/2] $. In particular, its maximum is given by $\max(q_{\gamma}(0),q_{\gamma}(\gamma)) = \max(2\sin_{\gamma}^{3},\sin_{\gamma})$.
\end{proof}

%\bibliographystyle{plain}
%\bibliographystyle{IEEEtran}
%\bibliography{bibliography,IFAC,ZelazoSnapshot}
\bibliography{fullbiblio}

\end{document}